\documentclass[aps,nofootinbib,prd,superscriptaddress,tightenlines,notitlepage,twocolumn,showpacs,floatfix]{revtex4-1}
\usepackage[colorlinks]{hyperref}
\usepackage{tabularx}
\usepackage{graphicx}
\usepackage{amssymb,amsmath,bm,tensor,braket}
\usepackage{xcolor}
\usepackage[varg]{txfonts}
\usepackage{enumerate}
\usepackage{mathtools}
\usepackage[capitalize]{cleveref}
\crefname{section}{Sec.}{Secs.}
\Crefname{section}{Section}{Sections}
\crefname{equation}{Eq.}{Eqs.}
\Crefname{equation}{Equation}{Equations}
\crefname{figure}{Fig.}{Figs.}
\Crefname{figure}{Figure}{Figures}
\crefname{subsection}{Sec.}{Secs.}
\usepackage[utf8]{inputenc}
\usepackage[normalem]{ulem}

\newcommand{\SUN}{\mathrm{M_{\odot}}}

\newcommand{\MPIfR}{\affiliation{Max-Planck-Institut f\"ur Radioastronomie, Auf
dem H\"ugel 69, D-53121 Bonn, Germany}}
\newcommand{\KIAA}{\affiliation{Kavli Institute for Astronomy and
Astrophysics, Peking University, Beijing 100871, China}}

\newcommand{\SOP}{\affiliation{School of Physics and State Key Laboratory of Nuclear Physics and Technology, Peking University, Beijing 100871, China}}

\newcommand{\CICQM}{\affiliation{Collaborative Innovation Center of 
                                Quantum Matter, Beijing, China}}	
\newcommand{\CHEP}{\affiliation{Center for High Energy Physics, 
                                Peking University, Beijing 100871, China}}
\newcommand{\BNU}{\affiliation{Department of Astronomy, Beijing
Normal University, Beijing 100875, China}}

\begin{document}

\title{Reduced-order surrogate models for scalar-tensor gravity in the 
strong field \\ and applications to binary pulsars and GW170817}

\date{\today}
\author{Junjie Zhao}\SOP
\author{Lijing Shao}\email{lshao@pku.edu.cn}\KIAA\MPIfR
%\author{Norbert Wex}\MPIfR
%\author{Alessandra Buonanno}\AEI\Maryland
%\author{Michael Kramer}\MPIfR\Manchester
%\author{Noah Sennett}\Maryland\AEI
\author{Zhoujian Cao}\BNU
\author{Bo-Qiang Ma}\email{mabq@pku.edu.cn}\SOP\CICQM\CHEP

\begin{abstract}
We investigate the scalar-tensor gravity of Damour and Esposito-Far\`ese (DEF),
which predicts non-trivial phenomena in the nonperturbative strong-field regime
for neutron stars (NSs). Instead of solving the modified
Tolman-Oppenheimer-Volkoff equations, we construct reduced-order surrogate
models, coded in the {\tt pySTGROM} package, to predict the relations of a NS
radius, mass, and effective scalar coupling to its central density. Our models
are accurate at $\sim1\%$ level and speed up large-scale calculations by two
orders of magnitude. As an application, we use {\tt pySTGROM} and Markov-chain
Monte Carlo techniques to constrain parameters in the DEF theory, with five
well-timed binary pulsars, the binary NS (BNS) inspiral GW170817, and a
hypothetical BNS inspiral in the Advanced LIGO and next-generation GW detectors.
In the future, as more binary pulsars and BNS mergers are detected, our
surrogate models will be helpful in constraining strong-field gravity with
essential speed and accuracy.
\end{abstract}

\maketitle

%---------------------------------------------------------------------
\section{Introduction}
\label{sec:intro}
%---------------------------------------------------------------------

The theory of general relativity (GR), proposed by Albert Einstein in 1915,
postulates that gravity is mediated only by a long-range, spin-2 tensor field,
$g_{\mu\nu}$~\cite{Einstein:1915ca}. For more than a century, tests of GR have
never stopped. The bulk of accurate experimental tests from, (i) the Solar
System~\cite{Will:2014kxa}, (ii) the high-precision timing of binary
pulsars~\cite{Stairs:2003eg, Wex:2014nva, Shao:2016ezh}, and (iii) the
gravitational-wave (GW) observations of coalescing binary black holes
(BBHs)~\cite{TheLIGOScientific:2016src, Abbott:2017vtc, LIGOScientific:2019fpa}
and binary neutron stars (BNSs)~\cite{TheLIGOScientific:2017qsa, GBM:2017lvd,
Abbott:2018lct}, have {\it all} been proven to be in line with GR.

However, there are various motivations to look for theories beyond
GR~\cite{Will:2018bme}. As one of the most natural alternatives, the
scalar-tensor gravity, in addition to $g_{\mu\nu}$, adds a long-range, spin-0
scalar field, $\varphi$. This theory was conceived originally by
Scherrer~\cite{Goenner:2012cq} and Jordan~\cite{Jordan:1949zz}. Viewpoints
similar to the one in the scalar-tensor gravity have also been sketched in the
Kaluza–Klein theory, the string theory, and the brane
theory~\cite{Fujii:2003pa}. The extra scalar degree of freedom is potentially
related to the dark energy, the inflation, and a possible unified theory of
quantum gravity.

From 1960s to the present time, a healthy version of scalar-tensor gravity has
played the most influential role (see Refs.~\cite{Fujii:2003pa, Will:2018bme}
for reviews). We now call it the Jordan–Fierz–Brans–Dicke (JFBD)
theory~\cite{Jordan:1949zz, Jordan:1959eg, Fierz:1956zz, Brans:1961sx}. Inspired
by the Mach's principle, the gravitational constant $G$ is promoted to a
time-varying dynamical field in the JFBD theory~\cite{Brans:1961sx}.  With the
additional scalar field coupled {\it nonminimally} to the Einstein-Hilbert
Lagrangian, the JFBD theory leads to a violation of the strong equivalence
principle (SEP)~\cite{Eardley:1975tv, Shao:2016ezh}.  In this paper, we focus on
a class of special mono-scalar-tensor gravity, formulated by Damour and
Esposito-Far\`ese (DEF)~\cite{Damour:1992we, Damour:1993hw, Damour:1996ke}.
Relative to the JFBD theory, the DEF theory extends the conformal coupling
function by including a quadratic term, which dictates the way matter couples to
the scalar field. Within a certain parameter space, it significantly modifies
the level where the SEP is violated for strongly self-gravitating
NSs~\cite{Damour:1993hw}.

The theory of scalar-tensor gravity has been extensively investigated in the
weak field, mainly from experiments in the Solar System~\cite{Will:2014kxa}.
The most stringent constraint comes from the Cassini
probe~\cite{Bertotti:2003rm}.  In the parametrized post-Newtonian (PPN)
framework~\cite{Will:2014kxa}, it is verified to a high precision $\sim 10^{-5}$
that the DEF theory is very close to GR in the weak-field
regime~\cite{Damour:2007uf}. In the nonperturbative strong-field regime,
\citet{Damour:1993hw} noticed a sudden strong activation of the scalar field for
NSs. This kind of intriguing feature, where the DEF theory can be reduced to GR
in the weak field while being significantly different in the strong field, has
caused enormous interests~\cite{Freire:2012mg, Antoniadis:2013pzd, Shao:2017gwu,
Anderson:2019eay}. The process that causes significant differences from GR is
referred as strong-field {\it scalarization}. It has been comprehensively
investigated for decades in terms of {\it spontaneous}
scalarization~\cite{Damour:1993hw, Damour:1996ke, EspositoFarese:2004cc,
Anderson:2016aoi, Anderson:2019eay, Anderson:2019hio}, as well as {\it induced}
and {\it dynamical} scalarizations~\cite{Barausse:2012da, Shibata:2013pra,
Sennett:2016rwa, Sennett:2017lcx, Shao:2017gwu}.

Different kinds of scalarization phenomena are defined in the following ways.
\begin{enumerate}
    \item Spontaneous scalarization occurs in an isolated and compact star
    whose compactness exceeds a critical value~\cite{Damour:1993hw,
    EspositoFarese:2004cc}. This phenomenon can be regarded as a phase
    transition~\cite{Damour:1996ke, Sennett:2017lcx}.
    \item Induced and dynamical scalarizations are discovered in
    numerical-relativity (NR) simulations of BNSs in the DEF theory. They
    hasten the plunge and merger phases relative to
    GR~\cite{Barausse:2012da}. The induced scalarization occurs in those
    binary systems, where one component of the binary has already
    spontaneously scalarized while the other has not in the early 
    inspiral~\cite{Barausse:2012da}.
    \item Dynamical scalarization corresponds to the binary systems, where
    none of the binary components can spontaneously scalarize in isolation, but 
    they get scalarized when the gravitational binding energy of the orbit 
    exceeds a critical value~\cite{Barausse:2012da}.
\end{enumerate}

In this paper, we pay particular attention to the strong-field region, where the
spontaneous-scalarization phenomena are significant. Compared with an
unscalarized system, the scalarized binary system brings the following manifest
changes: (i) an additional gravitational binding energy for the orbit, (ii) an
enhancement in the decay rate of binary's orbital period and the energy flux by
an extra dipolar radiation~\cite{Damour:2007uf}. As we know, dipolar radiation
corresponds to the $-1$ post-Newtonian (PN) correction.\footnote{We refer
$n\,$PN to $\mathcal{O}\left(v^{2n}/c^{2n}\right)$ corrections with respect to
the Newtonian order, where $v$ is the characteristic relative speed in the
binary. Here, we follow the convention in the GW phase, where the quadrupolar
radiation reaction is denoted as $0\,$PN.  Therefore, the dipolar radiation is
at $-1\,$PN order. } It enters at a lower order relative to the typical
quadrupolar radiation in GR. This means that, the smaller the relative speed,
the greater the relative radiating effect of the dipolar radiation. 

With the dominant radiating component being the dipolar emission at early time,
a binary system emits extra energy in addition to GR. In certain binary pulsar
systems, it is a powerful means to probe the strength of dipolar contribution
that can be caused by the spontaneous scalarization~\cite{Damour:1996ke,
Wex:2014nva}. With a small characteristic speed $\sim10^{-3}c$, the
gravitational radiation of binary pulsar systems could be dominated by the
dipolar emission.

A pulsar emits ratio signals like a lighthouse. In a binary pulsar system, the
spin period of the recycled pulsar is usually extremely stable. For several
years to decades, those periodic signals have been continuously monitored by
large radio telescopes on the Earth. It makes pulsar a {\it clock} that can
rival the best clocks for precision fundamental physics~\cite{Lorimer:2005misc,
Guo:2018rpw}. The accurate measurement technique, the so-called {\it pulsar
timing}, models the times of arrival (TOAs) of pulses emitted from the pulsar
and determines timing parameters to a high precision~\cite{Stairs:2003eg,
Lorimer:2005misc}. 

In order to accommodate alternative gravity theories, the parametrized
post-Keplerian (PPK) formalism was developed as a generic pulsar timing
model~\cite{Damour:1991rd}. 
A set of theory-independent Keplerian and post-Keplerian timing parameters
are determined with high precision in a fit of the timing model to the
TOAs~\cite{Damour:2007uf}. We
can obtain extremely precise physical parameters to describe those systems.
They can be used to place constraints on alternative gravity
theories~\cite{Stairs:2003eg, Wex:2014nva, Shao:2016ezh}. Binary pulsar is
currently one of the best available {\it strong-field testbeds} for testing
gravity~\cite{Damour:2007uf}.

Recently, GWs have started to compensate with binary pulsars in probing the
strong-field gravity. The first GW event of coalescing BNSs, GW170817, was
detected by the LIGO/Virgo Collaboration in August
2017~\cite{TheLIGOScientific:2017qsa}. GW170817 provides a powerful laboratory
in the highly dynamical strong field. The spacetime of BNSs is strongly curved
and highly dynamical in the vicinity of NSs in the late inspiral. If the DEF
theory correctly describes the gravity, GW phase evolution of BNSs is 
modified. For now, limited by the sensitivity of the LIGO/Virgo detectors below
tens of Hz, the precision to constrain the dipolar radiation from the short
duration of GW170817 is still less than binary pulsars~\cite{Shao:2017gwu}.

Observations of BNSs at lower frequency are beneficial in the
dipolar-radiation test. To increase the detector sensitivity, LIGO/Virgo have
once more upgraded their equipments, and recently started the observing run 3
(O3) on April 1, 2019.  Meanwhile, as the first kilometer-scale underground GW
detector, the Kamioka Gravitational Wave Detector
(KAGRA)~\cite{Kuroda:2010zzb,Aasi:2013wya} is likely to join O3 before the end
of 2019 as well~\cite{Akutsu:2018axf}. The next-generation ground-based GW
detectors, such as the Cosmic Explorer (CE) led by the United States, and the
Einstein Telescope (ET) led by the Europe~\cite{Hild:2010id}, will further
improve the sensitivity in the future. In particular, they extend the
sensitivity bands to be below 10\,Hz. At the time of the third-generation
detectors, we expect to discover more BNS merger events with higher
sensitivities and larger signal-to-noise ratios (SNRs).  These events are to put
more stringent limits on alternative gravity theories, with the DEF theory being
an important example.

In deriving constraints on the scalar-tensor gravity, the structure of NSs needs
to be solved~\cite{Damour:1993hw, Damour:1996ke}. Thus, the equation of state
(EOS) of NS matters plays a role. The EOS is used to infer a NS radius, mass,
and the scalar charge, by integrating the modified Tolman-Oppenheimer-Volkoff
(TOV) equations~\cite{Damour:1993hw,Damour:1996ke}. There are still large
uncertainties in the NS EOS. In this work we choose nine EOSs that are all
consistent with the maximum mass of NSs larger that $2\, \SUN$. Since more
observations are being made for pulsars at radio and X-ray wavelengths, and BNSs
with an increasing statistics, the uncertainty in the nuclear EOS is to be
reduced in the near future.

Given an EOS, with inputs from binary pulsars and the recent BNS observation, we
carry out advanced algorithms to constrain the DEF theory in a statistically
sound way. In particular, we employ Bayesian methods through Markov-chain Monte
Carlo (MCMC) simulations. Those simulations update posterior probability
distribution of parameters in the DEF theory by evaluating the likelihood
function millions of times. Every step requires the corresponding NS properties
(including mass, radius, and scalar charge), which are derived from the modified
TOV equations iteratively. Being computationally intensive, such studies have
been carried out already in Refs.~\cite{Shao:2017gwu, Anderson:2019eay}.

Being statistically sound, MCMC simulations lead to a large number of iterative
calculations however. Here we build a new model to reduce the computational
burden. Instead of solving the modified TOV equations iteratively and
repeatedly, we construct reduced-order surrogate models (ROMs) to predict NS
properties. There are two parameters characterizing the DEF theory, $\alpha_0$
and $\beta_0$ (see the next section). We explore a sufficiently large parameter
space for them in the regime of strong-field scalarization. With our surrogate
models, the process of obtaining NS properties is no longer an iterative
integration, but a linear algebraic operation. It costs a fixed amount of time,
much shorter than that in the previous method. In practice, for a given DEF
theory, we use the central matter density $\rho_c$ of a NS to predict its radius
$R$, mass $m_A$, and the effective scalar coupling $\alpha_A$. Our models are
numerically accurate at $\sim 1\%$ level for $\alpha_A$, and better than
$0.01\%$ for $m_A$ and $R$.  They accelerate the processes of parameter
estimation significantly everywhere in the parameter space we explore. With the
speedup of those models, one can perform MCMC simulations much more efficiently
yet still accurately. According to our performance test, they speed up
calculations at least one hundred times, compared with the previous method in
Ref.~\cite{Shao:2017gwu}. Various practical examples with binary pulsars and BNS
events are demonstrated in this paper.

The organization of the paper is as follows. In~\cref{sec:ssSTG}, we briefly
review the nonperturbative spontaneous-scalarization phenomena for isolated NSs.
The additional dipolar radiation and the modification of mass-radius relations
for different EOSs in the scalar-tensor gravity will be discussed.
\Cref{sec:ROM} analyzes the difficulties in solving the modified TOV equations
with large-scale calculations. We develop a better numerical method, and code it
streamlinedly in the {\tt pySTGROM} package. We make it public for an easy use
for the community.\footnote{\url{https://github.com/BenjaminDbb/pySTGROM}}
In~\cref{sec:PSRAndGW}, with the speedup from {\tt pySTGROM}, we stringently
constrain the DEF theory by combining the dipolar-radiation limits from
observations of five NS-white dwarf (WD) systems, and that of GW170817, which
also includes a modified mass-radius relation. Our results are in good agreement
with that from \citet{Shao:2017gwu}. We also forecast  constraints involving a
hypothetical BNS event, to be detected by the Advanced LIGO at its design
sensitivity~\cite{TheLIGOScientific:2014jea}, and next-generation ground-based
GW detectors. Finally, the main conclusions and discussions are given
in~\cref{sec:conc}.

%---------------------------------------------------------------------
\section{Spontaneous scalarization in the DEF gravity}
\label{sec:ssSTG}
%---------------------------------------------------------------------

In our study, we concentrate on the DEF theory. It is defined by the following
action in the {\it Einstein frame}~\cite{Damour:1993hw,Damour:1996ke},
\begin{align}
  \label{eqn:action}
   S &= \frac{c^4}{16 \pi G_*} \int \frac{{\rm d}^4 x}{c} \sqrt{-g_*}
   \left[ R_* - 2 g^{\mu\nu}_* \partial_\mu \varphi \partial_\nu \varphi
   - V(\varphi) \right] \nonumber\\
   & \qquad + S_m \left[ \psi_m; A^2(\varphi) g^*_{\mu\nu} \right] \,.
\end{align}
In~\cref{eqn:action}, $g_* \equiv {\rm det} \, g_{\mu \nu}^* $ denotes the
determinant of the Einstein metric $g_{\mu \nu}^*$, $R_*$ is the Ricci curvature
scalar of $g_{\mu \nu}^*$, $G_*$ is the bare gravitational coupling constant,
$\varphi$ is the dynamical scalar field that is added to GR, $\psi_m$ describes
any matter fields, and $A(\varphi)$ is the conformal coupling factor that
determines how $\varphi$ couples to $\psi_m$ in the Einstein frame.

The potential of the scalar field $V(\varphi)$ can be neglected for a slowly
varying $\varphi$ compared with the typical scale of the system. Therefore, in
our study we set $V(\varphi)=0$. Alternatively,
Refs.~\cite{Alsing:2011er,Ramazanoglu:2016kul} considered the effects of a
massive scalar field, via $V(\varphi) \approx 2 m_{\varphi} \varphi^2$.  The
field equations of the DEF theory can be derived by varying the
action~(\ref{eqn:action}). They are~\cite{Damour:1993hw, Damour:1996ke},
\begin{align}
  R^*_{\mu\nu} &= 2 \partial_\mu \varphi \partial_\nu \varphi
  + \frac{8\pi G_*}{c^4} \left(T^*_{\mu\nu}
  - \frac{1}{2} T^* g^*_{\mu\nu} \right) \, , 
  \label{eqn:field:metric} \\
  \square_{g^*} \varphi &= -\frac{4\pi G_*}{c^4} \alpha(\varphi) T_* \,,
  \label{eqn:field:scalar}
\end{align}
where $T_*^{\mu\nu} \equiv 2c \left(-g_*\right)^{-1/2} \delta S_m / \delta
g^*_{\mu\nu} $ denotes the matter stress-energy tensor and,
\begin{equation}
  \alpha(\varphi) \equiv \frac{\partial \ln A(\varphi)}{\partial \varphi} \, .
  \label{eqn:alphaPhi}
\end{equation}
As~\cref{eqn:field:scalar} shows, $\alpha(\varphi)$ measures the field-dependent
coupling strength between the scalar field $\varphi$ and the trace of the
energy-momentum tensor of matter fields, $T_* \equiv g^*_{\mu \nu} T_*^{\mu
\nu}$.

In the JFBD theory, $\ln A(\varphi)$ was chosen to be linear in $\varphi$, i.e.,
$\ln A(\varphi) = \alpha_0 \varphi$. It is extended to the polynomial form up to
the quadratic order in the DEF theory~\cite{Damour:1993hw},
\begin{equation}
  \ln A(\varphi)= \frac{1}{2} \beta_{0}\varphi^{2} \, ,
  \label{eqn:lnA}
\end{equation}
with $\alpha(\varphi) \equiv \partial \ln A(\varphi) / \partial \varphi =
\beta_0 \varphi$. We denote $\alpha_0 = \beta_0 \varphi_0$ with $\varphi_0$
being the asymptotic (cosmological background) scalar field value of $\varphi$
at infinity. Therefore, there are only two extra parameters, $\alpha_0$ and
$\beta_0$ (or equivalently, $\varphi_0$ and $\beta_0$), to describe a DEF theory
uniquely. In GR, we have $\alpha_0 = \beta_0 = 0$.

Within the PPN framework in the weak field, Solar-System experiments can be used
to narrow down the parameter space in the DEF theory.  However, generally only
the $\alpha_0$ or the combination $\beta_0\alpha_0^2$ can be constrained~(see
Refs.~\cite{Damour:2007uf,Will:2014kxa}). 

For NSs, the aforementioned nonperturbative scalarization phenomena happen
when~\cite{Damour:1993hw, Barausse:2012da},
\begin{equation}
  \beta_0 \equiv \left. \frac{\partial^2 \ln A(\varphi)}{\partial \varphi^2} \right|_{\varphi = \varphi_0}
    \lesssim -4 \, .
  \label{eqn:beta0}
\end{equation}
From Fig.~1 in Ref.~\cite{Shao:2017gwu}, it is clear that with certain condition
a negative $\beta_0$ can trigger an instability in the scalar
field~\cite{Sennett:2017lcx}. It describes the strength of nonperturbative
phenomena. The more negative for $\beta_0$ from the critical value $-4.0$, the
more manifest the spontaneous scalarization in the strong-field regime.

In the strong field, the {\it effective scalar coupling} for a NS $A$, 
\begin{equation}
  \alpha_A \equiv  \left. \frac{\partial \ln m_A(\varphi)}{\partial \varphi} \right|_{\varphi=\varphi_0} \, ,
  \label{eqn:alphaA}
\end{equation}
measures the ``sensitivity'' of the coupling between the NS mass and variations
in the background scalar field $\varphi_0$. It appears directly in the Keplerian
binding energy between two stars, $A$ and $B$,
\begin{equation}
  V_{\rm int} = - \frac{G_* \, m_A m_B }{r_{AB}} \left(1 + \alpha_A \alpha_B \right).
  \label{eqn:binding}
\end{equation}
Besides the gravitational attraction in GR, the effective scalar coupling,
$\alpha_A$, brings an additional scalar interaction. It also affects the
strength of the orbital period decay of binaries~\cite{Damour:1996ke}.

In the following, we investigate the dipolar contribution to the orbital decay
from the scalar field, $\dot P_b^{\rm dipole}$, and the quadrupolar contribution
from the tensor field, $\dot P_b^{\rm quad}$. They read~\cite{Peters:1963ux,
Damour:1993hw}
\begin{equation}
  \dot P_b^{\rm dipole} = - \frac{2 \pi  G_*}{c^3} g(e)
  \left(\frac{2\pi}{P_b}\right)
  \frac{m_p m_c}{m_p + m_c}
  \left(\alpha_p - \alpha_c\right)^2 \,,
  \label{eqn:pbdot:dipole}
\end{equation}
\begin{equation}
  \dot P_b^{\rm quad} = -\frac{192\pi G_*^{5/3}}{5c^5}
  f(e) \left(\frac{2\pi}{P_b}\right)^{5/3}
  \frac{m_p m_c}{\left(m_p + m_c\right)^{1/3}} \,,
  \label{eqn:pbdot:quad}
\end{equation}
where $P_b$ is the orbital period, ``$p$'' and ``$c$'' denote the pulsar and its
companion respectively. Subdominant contributions to $\dot P_b^{\rm quad}$, that
are totally negligible to our study, can be found in Eq.~(6.52d) in
Ref.~\cite{Damour:1992we}. In above equations, $g(e)$ and $f(e)$ are functions
of the orbital eccentricity $e$,
\begin{align}
  g(e) &\equiv \left(1 - e^2\right)^{-5/2} \left(1 + \frac{e^2}{2}\right) \,, \\
  f(e) &\equiv \left(
    1-e^2\right)^{-7/2}
    \left(1 + \frac{73}{24} e^2 + \frac{37}{96} e^4\right)  \,.
\end{align}
In~\cref{eqn:pbdot:dipole,eqn:pbdot:quad}, $G_N = G_* \left(1 + \alpha_0^2
\right)$ denotes the Newtonian gravitational coupling in the weak
field~\cite{Damour:2007uf}.

The effective scalar coupling equals to zero for BHs due to the no-hair
theorem~\cite{TheLIGOScientific:2016src,Berti:2015itd}, and approaches to
$\alpha_0$ for weak-field WDs in the DEF theory. Therefore, the influences from
dipolar radiation on the orbital period decay occur dominantly in NS-WD and
NS-BH binaries, as well as in asymmetric NS-NS binaries.\footnote{In the BH-BH
systems, scalar charges of binaries both equal to zero. They hardly contribute
to the orbital decay. But in the most general scalar-tensor theory, the
Horndeski gravity~\cite{Horndeski:1974wa}, 
such systems are able to have scalar hairs which, depending on the details of
the theory, might be the case only for certain mass ranges.} Since the
dipolar contribution relates to
$-1\,$PN contribution in the GW phase, instructively, a precise bound on the
gravitational dipole emission can be derived with the early inspiral stage of
relevant GW events.

For a comprehensive study on NS spontaneous scalarization in the strong field,
we here turn our attention to the nuclear EOS and the other NS properties.
Usually, the NS matter can be treated as a perfect fluid. In GR, given an EOS,
we can solve the classical TOV equations~\cite{Tolman:1939jz,Oppenheimer:1939ne}
for NSs. The NS radius, $R$, and mass, $m_A$, can be derived when the central
matter density, $\rho_c$, and the EOS are given. In the DEF theory, we involve
the additional scalar field $\varphi$. Correspondingly, the modified TOV
equations [Eq.~(7) in Ref.~\cite{Damour:1993hw} and Eqs.~(3.6a)~to~(3.6f) in
Ref.~\cite{Damour:1996ke}] should be used. In the {\it Jordan
frame},\footnote{The Jordan frame, also known as the physical frame, equivalents
to the Einstein frame by a conformal transformation with redefinitions of the
metric and the scalar field~\cite{Dicke:1961gz,Weyl:1993kh}.} physical
quantities, $R$, $m_A$, and $\alpha_A$, can be obtained via integrating from
$\rho_c$ and the value of the scalar field at the center of a NS, $\varphi_c$.

\begin{figure}
  \includegraphics[width=8.66cm]{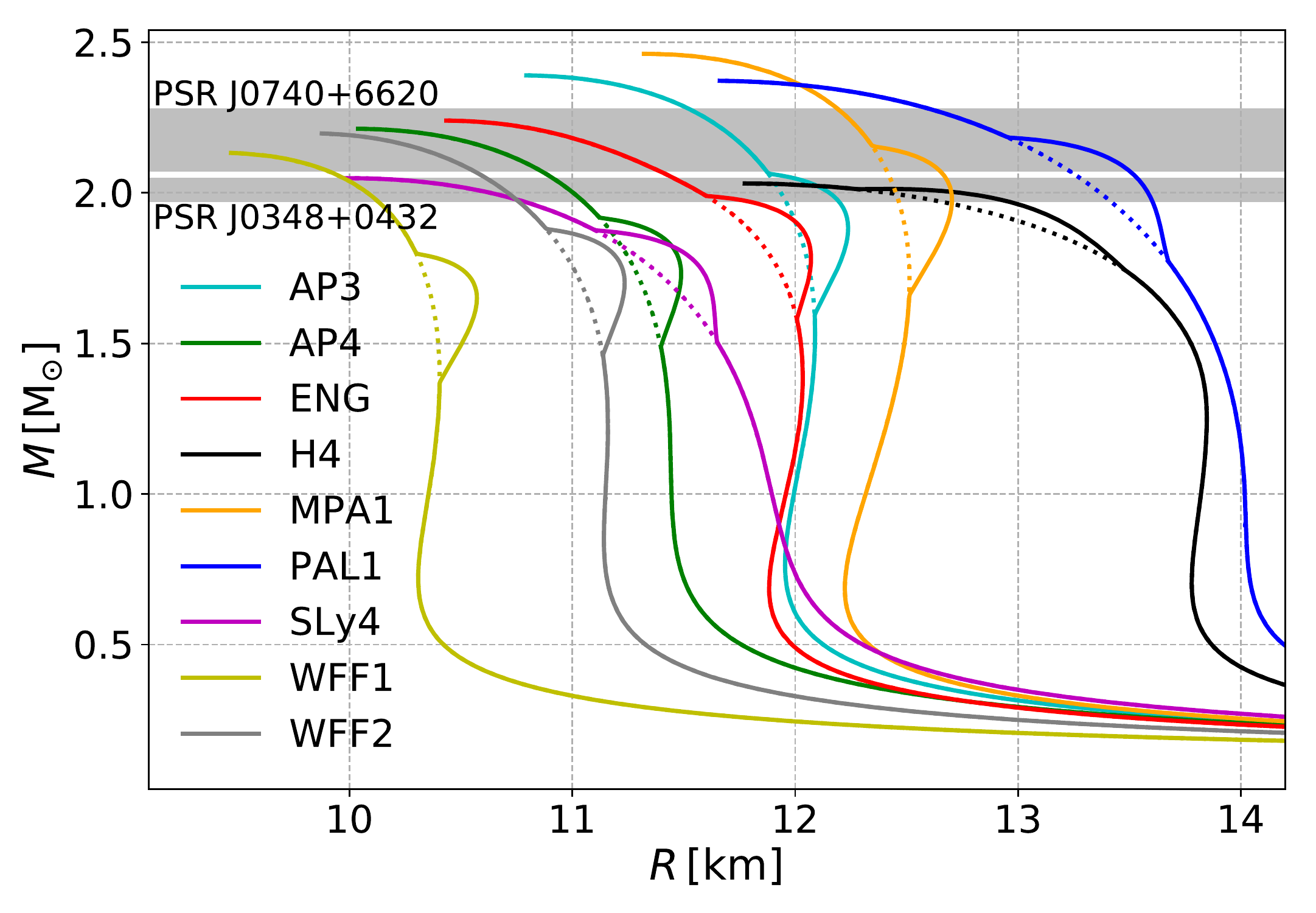}
\caption{(color online) The mass-radius relations of NSs for different EOSs.  We
adopt nine EOSs whose corresponding maximum masses for NSs are larger than
$2\,\SUN$. The relations of mass and radius are derived from GR (dotted lines),
and from the DEF theory with $\log_{10} |\alpha_0| = -5.0$ and $\beta_0 = -4.5$
(solid lines). The precise mass constraints (indicated with their 1-$\sigma$
uncertainties) from PSRs J0740$+$6620~\cite{Cromartie:2019kug} and
J0348$+$0432~\cite{Antoniadis:2013pzd} are depicted in grey. In the figure, the
curves from GR and the DEF theory overlap largely, except the ``bumps'' that
result from the nonperturbative spontaneous scalarization. In the region of
bumps, the DEF theory shows a larger radius for a same NS mass relative to GR.
\label{fig:MRRelation}}
\end{figure}

We show an example of mass-radius relation for NSs in the DEF theory with
$\log_{10} |\alpha_0| = -5.0$ and $\beta_0 = -4.5$
in~\cref{fig:MRRelation}~\cite{Shao:2019gjj}. Two massive pulsars, PSRs
J0740$+$6620~\cite{Cromartie:2019kug} and
J0348$+$0432~\cite{Antoniadis:2013pzd}, show that the maximum mass of NSs is
most likely above $2 \, \SUN$. Therefore, we choose the EOSs that satisfy this
condition in our study. In total, we use nine EOSs. From \cref{fig:MRRelation}
we can see that,
\begin{enumerate}
    \item the mass-radius relation in the DEF theory is very close to
    that of GR in the majority of mass range;
    \item there are ``bumps'' that correspond to the spontaneous
    scalarization phenomena and, when compared with GR, the bump for each EOS
    predicts a larger radius for a same NS mass;
    \item in GR, the EOSs, {\sf SLy4} and {\sf H4}, are getting more and more
    in tension with pulsar mass measurements, because their maximum masses
    hardly reach
    the 1-$\sigma$ lower limit of PSR J0740$+$6620~\cite{Cromartie:2019kug}.
\end{enumerate}
It is worthy to note that, the EOSs {\sf H4} and {\sf PAL1} predict relatively
larger radii for NSs. They are starting to be disfavored by the LIGO/Virgo's
observation of GW170817~\cite{Abbott:2018exr, Abbott:2018wiz}.

%---------------------------------------------------------------------
\section{The reduced-order model}
\label{sec:ROM}
%---------------------------------------------------------------------

As discussed in the previous section, the initial values for the TOV integrator
include the NS center matter density $\rho_c$, and the center value of the
scalar field $\varphi_c$. The macroscopic quantities of NSs, namely $R$, $m_A$,
and $\alpha_A$, can be obtained from integrating the modified TOV equations.
From the viewpoint of the DEF gravity, instead of $\varphi_c$, it is more
convenient to have the scalar field at infinity, $\varphi_0$, as an initial
value. There is correspondence between $\varphi_0$ and $\varphi_c$. Thus, to
obtain a desired $\varphi_0$, we need to find the value of $\varphi_c$. To solve
such a boundary value problem, the shooting method turns out to be a practical
way~\cite{Shao:2017gwu}. Nevertheless, sometimes it leads to a large number of
iterations, especially around the region of spontaneous scalarization.
Therefore, it is not efficient to perform large-scale calculations, such as the
parameter estimation with the MCMC approach, based on the shooting method. It is
helpful to find a faster and more efficient, but still accurate, surrogate
algorithm for a better performance in solving the NS structure in the DEF
theory.

Before discussing the surrogate algorithm for the DEF theory, we turn to GW data
analysis for idea. A similar issue, that is equally computationally challenging
and difficult, arises when considering GW data analysis~\cite{Veitch:2014wba}.
In the case of compact binary coalescences, the matched-filtering technique is
used. It involves matching the data against a set of template waveforms (see
e.g. Ref.~\cite{Bohe:2016gbl}), to infer a potential astrophysical signal. The
analysis could lead to a large computational cost. Therefore, effective
approaches, the surrogate models, were built.

Distinct examples are developed by~\citet{Purrer:2014fza,Purrer:2015tud}
and~\citet{Field:2013cfa,Field:2011mf}. This kind of model is constructed by a
set of highly accurate sparse waveforms. It brings a new waveform model to
provide fast and accurately compressed approximations. These accurate surrogate
models are built without sacrificing the underlying accuracy for data analysis.
Recently, several such methods have been proposed in the literature. Based on
the singular value decomposition (SVD) method, time-domain inspiral surrogate
waveforms are generated in Ref.~\cite{Purrer:2014fza}. Another popular
construction method is the greedy reduced basis method, which is usually
combined with the empirical interpolation method
(EIM)~\cite{doi:10.1137/090766498,maday:hal-00174797}. It has been
applied to GW waveforms~\cite{Field:2013cfa} and BH
ringdown~\cite{Caudill:2011kv}.

Inspired by the above optimization for large-scale calculations in the GW data
analysis, we adopt the greedy reduced basis method to the NS structure in the
DEF theory. In our surrogate models, for each EOS, the NS properties are
accurately and rapidly inferred with given NS central matter density $\rho_c$.
In the following subsections, we briefly introduce the processes of general
ROM, and describe in detail how to specialize this model to the DEF theory.
Finally, we assess our ROMs' accuracy, and find that the final results are
consistent with our expectation. Our ROMs will be a helpful tool in generating
NS properties with essential speed and accuracy for future studies.

%---------------------------------------------------------------------
\subsection{A brief technical introduction to ROM}
\label{ssec:introROM}
%---------------------------------------------------------------------

The theoretical aspects and the processes of building the ROM have been
discussed extensively in Fig.~1 and Appendices~A~to~D in
Ref.~\cite{Field:2013cfa}. We here give a brief overview.

Let us use the GW data analysis as a prototype, and denote a curve $h(t;
\boldsymbol{\lambda})$, representing a waveform with parameters
$\boldsymbol{\lambda}$ (for GW waveforms, parameters in $\boldsymbol{\lambda}$
include the masses, the spins, and so on). In order to generate a ROM for $h(t;
\boldsymbol{\lambda})$, we need the data of $h(t)$, produced with a set of given
parameters $\left\{\boldsymbol{\lambda}_i\right\}$ on a grid.  We denote the
training space ${\mathbf{V} \equiv \left\{h(t; \boldsymbol{\lambda}_i)
\right\}}$. With the reduced basis (RB) method, we select a certain number of
bases derived from the training space $\mathbf{V}$. Those bases are regarded as
a set of representative bases for the remaining waveforms. Instead of choosing
$m$ existing $\{\boldsymbol{\lambda}_i\}$ and their corresponding space $\{h(t;
\boldsymbol{\lambda}_i) \}$, we seek a more complete set of bases that are as
independent as possible from each other.

The greedy algorithm is currently one of the methods to select $m$ orthonormal
RBs (see Ref.~\cite{doi:10.1137/100795772} and  Appendix. A in
Ref.~\cite{Field:2013cfa} for details).  The corresponding chosen space is
$\mathbf{RV} = \left\{e_i\right\}_{i=1}^m$.  Actually, the process of generating
orthonormal RBs is mentioned as iterated Gram-Schmidt orthogonalization
algorithm with greedy
selection~\cite{Ruhe:1983,Giraud:2005,Hoffmann1989,doi:10.1109/MCSE.2018.042781323}.
Firstly, a seed, which can be any waveform in $\mathbf{V}$, is chosen as the
starting RB $(i=0)$. Then, we define the maximum projection error,
\begin{equation}
  \sigma_{i} \equiv \max _{h \in \mathbf{V}}\left\|h(\cdot ; \boldsymbol{\lambda})
                -\mathcal{P}_{i} h(\cdot ; \boldsymbol{\lambda})\right\|^{2}\,,
  \label{eqn:maxProjErr}
\end{equation}
and a user-specified tolerable error bound, $\Sigma$, to infer the next RB
iteratively in the way that is explained here. In~\cref{eqn:maxProjErr},
$\mathcal{P}_{i}$ describes the projection of $h(t;\boldsymbol{\lambda})$ onto 
the span of the first $i$ RBs. The waveform corresponding to the maximum $\sigma_i$
is chosen as the next $(i+1)$-th RB, $e_{i+1}(t)$, after orthogonalized by the
iterated Gram-Schmidt algorithm. In practice, $\Sigma$ is set as a threshold to
terminate iterations of the greedy selection. When $\sigma_{m-1} \lesssim
\Sigma$, the desired orthonormal $m$ bases are complete.

With RBs obtained above, every waveform in the training space is well
approximated by an expansion of the form,
\begin{equation}
  \label{eqn:expansion}
  h(t; \boldsymbol{\lambda}) \approx 
\sum_{i=1}^{m} c_{i}(\boldsymbol{\lambda}) \, e_{i}(t) \approx 
\sum_{i=1}^{m} \braket{h(\cdot ; \boldsymbol{\lambda}), e_{i}(\cdot)} \, e_{i}(t) \,.
\end{equation}
In~\cref{eqn:expansion}, $c_{i}(\boldsymbol{\lambda})$ is the corresponding
coefficient of the $i$-th RB. It is calculated by a special ``inner product'' in
the space $\mathbf{RV}$. We collect such coefficients derived from $\mathbf{V}$
as the greedy data for the ROM. After that, we perform the EIM algorithm (see
Appendix~B in Ref.~\cite{Field:2013cfa} for details) to identify $m \,$
time-samples, which we call the empirical nodes. An interpolation can be built
to accurately reconstruct any fiducial waveform by the RBs. Finally, at each
empirical node, we perform a fit to the parameter space,
$\{\boldsymbol{\lambda}_i\}$, and finish the construction of the
ROM.\footnote{Particularly, a 2-dimensional 5th spline interpolating fit is used in
our study. A different 2-dimensional fitting method works without practical
difference.} For convenience, the fitted data are saved in a model file.

Now, we assume that a waveform, parametrized by an unknown set of parameters
$\boldsymbol{\lambda}_\star$, is required in the calculation. The parameters
$\boldsymbol{\lambda}_\star$ are different from $\boldsymbol{\lambda}_i$, but
within the boundary ranges of $\left\{\boldsymbol{\lambda}_i \right\}$. We can
use the ROM and the fitted data on each empirical node to get a complete
waveform as a function of $t$, $h(t; \boldsymbol{\lambda}_\star)$. The value of
$h(t_\star; \boldsymbol{\lambda}_\star)$ is derived for a given $t_\star$ in a
fast and accurate manner.

%---------------------------------------------------------------------
\subsection{Constructing ROMs for the DEF gravity}
\label{ssec:conROM}
%--------------------------------------------------------------------

In this subsection, we specialize the greedy reduced basis method to build the
ROMs for the DEF theory. The detailed approach is explained according to the
steps outlined above. Because the idea of building ROMs is ``borrowed'' from the
GW waveform studies, we use the word ``waveform'' to represent the desired
functional forms. The parameters $\boldsymbol{\lambda}$ are specialized as 
$(\alpha_0, \beta_0)$.

\begin{figure*}
  \includegraphics[width=17.9cm]{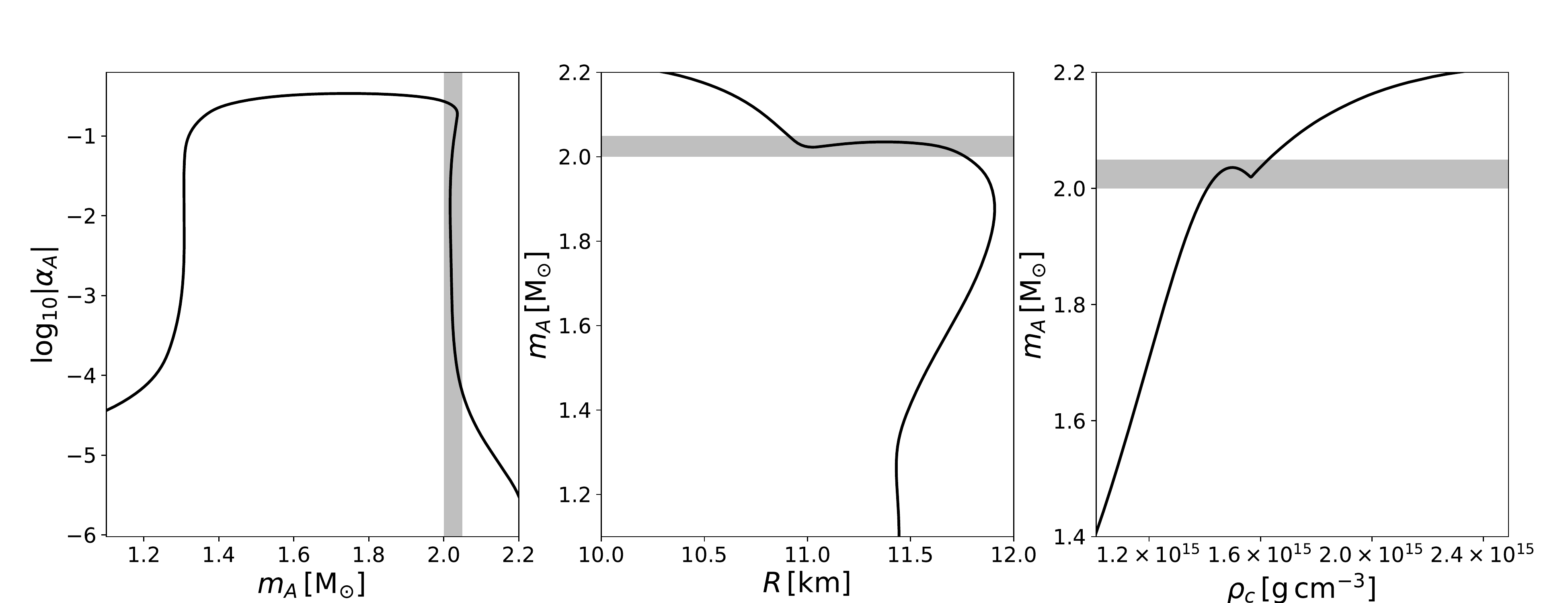}
\caption{{\it Pathological} phenomena occur when integrating the modified TOV
equations. The curves represent different relations between NS properties.  The
calculation assumes the DEF parameters, $\log_{10} |\alpha_0|=-5.3$ and
$\beta_0=-4.8$. We use the EOS {\sf AP4}. All grey bands denote the mass range
$m_A \in \left[2.0\,\SUN, 2.05\,\SUN\right]$ for a NS. The left panel shows the
relation between $\log_{10} |\alpha_A|$ and $m_A$. Notably, the multivalued
relation between $\alpha_A$ and $m_A$ arises in the grey band.  The middle panel
shows a similar pathology for building the ROM. In the right panel, an excessive
center matter density (about $\rho_c \gtrsim 1.5 \times 10^{15} \, {\rm g
\,cm^{-3}}$) of the NS leads to a collapse, and form a BH rather than a NS.
These pathologies are caused by the gravitational instability when $\rho_c$
exceeds a critical value. \label{fig:odrerr}}
\end{figure*}

\begin{figure}
  \includegraphics[width=8.66cm]{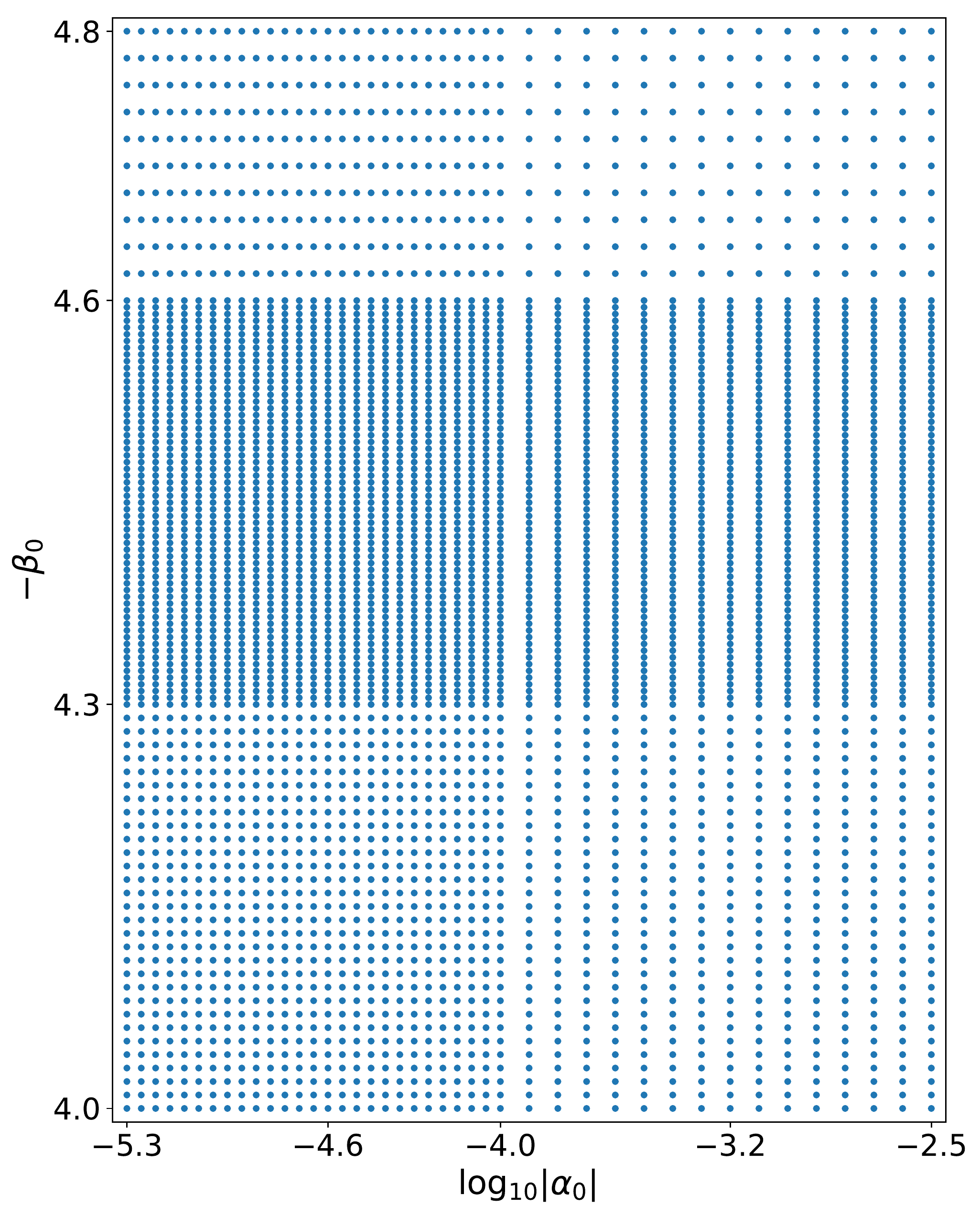}
\caption{(color online) An {\it uneven} grid in the parameter space $(\log_{10}
|\alpha_0|, - \beta_0)$ is used in building the ROMs of the DEF theory. We
generate a set of 42 $\times$ 101 = 4242 ``waveforms'' as the training data.}
  \label{fig:griddat}
\end{figure}

Ideally we want to get the NS properties, $R$ and $\alpha_A$, as a function of
the mass parameter, $m_A$. But in some parameter space {\it pathological}
behaviors might happen, while simply integrating the modified TOV equations.  We
use the DEF theory with $\log_{10} |\alpha_0|=-5.3$ and $\beta_0=-4.8$ for the
EOS {\sf AP4} as an example. We illustrate the different relations of NS
effective scalar coupling $\alpha_A$, radius $R$, and center matter density
$\rho_c$ in~\cref{fig:odrerr}. In the left panel, the effective scalar coupling
can obtain a value significantly larger than $0.1$ within the NS mass interval
$m_A \in \left[1.4\, \SUN, 2.0\,\SUN \right]$. This strength of spontaneous
scalarization in general decreases monotonically and rapidly when $m_A \gtrsim
2.0\, \SUN$. But, the multivalued relations between $\alpha_A$ and $m_A$ arise
when $\beta_0 \lesssim -4.6$, namely, multiple values for $\alpha_A$ are
possible for a given $m_A$. The curve in the middle panel shows the mass-radius
relation where the ``bump'' happens (see \cref{fig:MRRelation} for a similar
behavior). There exists the multivalued relation as well. In the bump region, an
$m_A$ can correspond to multiple values of $R$ within the interval $m_A \in
\left[2.0 \, \SUN, 2.05 \, \SUN \right]$, as indicated in grey. Similarly, in
the right panel, the mass of NS has a sudden drop when $\rho_c$ exceeds a
certain value in the grey band.

In~\cref{fig:odrerr}, in the interval $m_A \in \left[2.0 \, \SUN, 2.05 \, \SUN
\right]$, a fixed NS mass $m_A$ can correspond to multiple values for $\alpha_A$
and $R$. The existences of those phenomena have also been confirmed in the other
EOSs, when $\beta_0$ is below a critical value. For the EOS {\sf AP4}, the
critical value approximates to $-4.6$. Combining this observation and the
relation of $m_A$ and $\rho_c$ in the right panel of \cref{fig:odrerr}, we have
become fully aware where the pathologies come from. It is due to the excessive
density at the center (about $\gtrsim 1.5 \times 10^{15} \, {\rm g \,cm^{-3}}$
for the EOS {\sf AP4}), which causes the NS to further collapse into a BH. In
order to avoid dealing with those multivalued relations in the ROMs, we promote
the implicit parameter $\rho_c$, to an independent variable which corresponds to 
the ``time'' $t$ in~\cref{ssec:introROM}. Therefore, we
finally decide to construct three independent ROMs for the DEF theory,
$m_A\left(\rho_c\right)$, $R\left(\rho_c\right)$, and
$\alpha_A\left(\rho_c\right)$.\footnote{Actually, we use 
$\ln \left| \alpha_A \right| $ instead of $\alpha_A$ in the ROM for a better 
performance.} Those relations are all single-valued. The NS
properties, $(R, m_A, \alpha_A)$, are expected to be derived with a given
parameter $\rho_c$ {\it injectively} in the ROMs. 

For different EOSs, the ranges of $\rho_c$ are different. We
choose the range of $\rho_c$ to have $m_A \in (1\,{\rm M}_\odot, m_A^{\rm
max})$ where the maximum NS mass $m_A^{\rm max}$ is EOS-dependent. The
spacing $\Delta \rho_c$ is selected according to the steepness of the
scalarization in different regions where we choose more points in rapidly
changing parameter space. After building the ROMs, we can have results for
all $\rho_c$ in its range directly from our ROMs.

In a short summary, we build three ROMs, $m_A(\rho_c)$, $R(\rho_c)$, and
$\alpha_A(\rho_c)$. Though with one more implicit parameter $\rho_c$. This
choice has avoided the pathologies if we had built two ROMs, $R(m_A)$ and
$\alpha_A(m_A)$.

For the sake of completeness, we should generate all training waveforms of
interest at locations lying in the parameter space grid,
\begin{equation}
  \mathbf{V} \equiv \mathcal{A} \times \mathcal{B}\,.
  \label{eqn:VAB} 
\end{equation}
In~\cref{eqn:VAB}, $\mathcal{A}$ and $\mathcal{B}$ are 1-dimensional sets
covering the desired parameter ranges. They are representing the parameters in
the DEF theory, $\alpha_0$ and $\beta_0$, respectively. The spacing in the
parameters of $\log_{10}|{\alpha_{0}}|$ and $-\beta_{0}$ is chosen in an uneven
way, shown in~\cref{fig:griddat}. We concentrate more samples near the rapidly
changing domain in the greedy data for each empirical node.  Particularly, we
choose $\mathcal{A}$ to cover the range of $\log_{10} |\alpha_{0}| \in [-5.3, \,
-2.5]$ with 42 nodes. For $\beta_{0}$, 101 nodes are chosen to cover the
interval $-\beta_0 \in [4.0, 4.8]$. Totally, we have involved $42 \times 101 =
4242$ sparse waveforms to form the training space $\mathbf{V}$ for constructing
the ROMs. For the boundary values in building the ROMs, the particular value
$\alpha_0 \approx 10^{-2.5}$ approximately equals to the upper limit given by
the Cassini spacecraft~\cite{Bertotti:2003rm,Damour:2007uf}, and the critical
value $\beta_0 \lesssim -4.0$ indicates places where the spontaneous
scalarization happens in the DEF theory~\cite{Damour:1993hw, Barausse:2012da}. 
These choice will also serve as our priors in the parameter estimation that we 
will discuss in~\cref{sec:PSRAndGW}.

The greedy selection of RBs from the training space $\mathbf{V}$ is iteratively
performed to generate the next RB until the desired accuracy is achieved, namely
$\sigma_i \lesssim \Sigma$. In practice, $\Sigma$ is set to terminate the
iteration of the greedy selection. In our ROMs, we choose $\Sigma = 10^{-7}$ for
$R$ and $m_A$, and $\Sigma = 10^{-5}$ for $\alpha_A$. 

\begin{figure}
  \includegraphics[width=8.66cm]{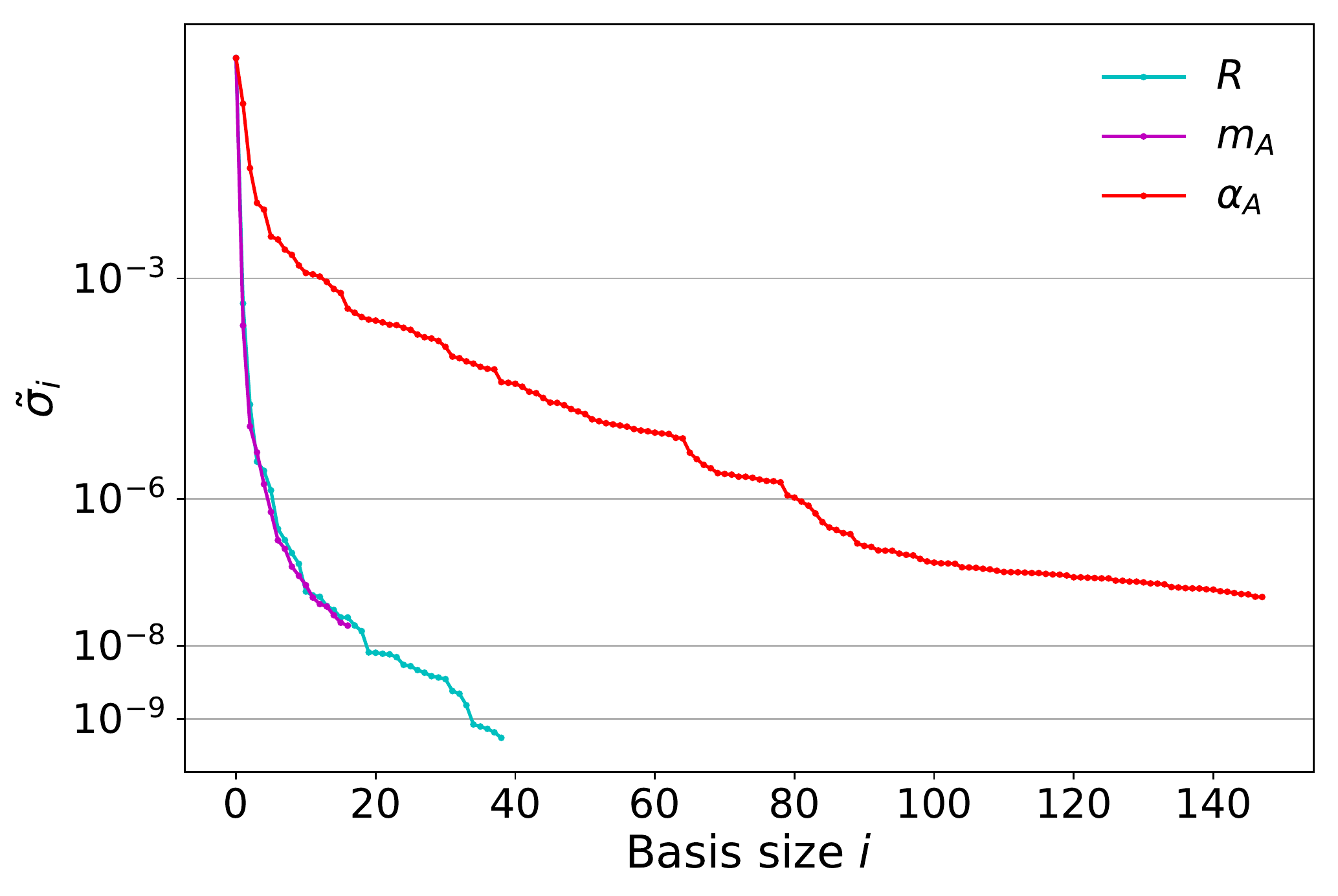}
\caption{(color online) The relative maximum projection error,
$\tilde{\sigma}_i$, in building the ROMs for the EOS {\sf AP4}. For the ROMs of
$R$ and $m_A$, we set $\Sigma = 10^{-7}$, and for the ROM of $\alpha_A$, we set
$\Sigma=10^{-5}$. As the orthonormal basis size increases in the greedy
selection, the $\tilde{{\sigma}_i}$'s of $R$ and $m_A$ decrease rapidly to $\sim
\,10^{-8}$ with a basis size of 20. In contrast, for $\alpha_A$, the error falls
off slowly and smoothly to the level of $10^{-7}$ for a few hundreds of basis
size. \label{fig:projErr}}
\end{figure}

In the iterative process, the relative projection errors, $\tilde{\sigma}_i
\equiv \sigma_i / \sigma_0$, are recorded for convenience. We illustrate
$\tilde{\sigma}_i$ as a function of the basis size in~\cref{fig:projErr}. In the
figure, the most notable features are the decreasing speed for
$\tilde{\sigma}_i$ with an increasing basis size. As the basis size increases,
the $\tilde{\sigma}_i$'s of $R$ and $m_A$ quickly arrive at the level $\lesssim
10^{-8}$ within the basis size of 20. Particularly, the ROM of $m_A(\rho_c)$ has
the fastest decline. In contrast, $\tilde{\sigma}_i$ falls off smoothly to
$\Sigma \sim 10^{-5}$ for about 150 steps in building the ROM of $\alpha_A$.
This means that, more RBs are needed to ensure the accuracy of $\alpha_A$.
Actually, considering the tolerable error involved by the shooting method when
integrating the modified TOV equations, which is about $\sim \, 1\%$, we find
that the precision loss of the above processes in constructing ROMs almost
negligible. This conclusion is verified when assessing the accuracy of the
ROMs in \cref{ssec:assessROM}.

Finally, we follow the steps in~\cref{ssec:introROM} to obtain RBs with the EIM
algorithm, and build the entire ROMs. Those ROMs can generate NS properties,
$R$, $m_A$, and $\alpha_A$, efficiently and rapidly as a function of $\rho_c$.
We tested the ROMs with randomly generated parameters, and found that the time 
for one solve of the modified TOV equations improves from $\sim 300\,$milliseconds 
to $\sim1\,$millisecond on our {\sf Intel Xeon E5} computers.
We implement those three ROMs for the DEF theory in the {\tt pySTGROM} package.
As is shown with applications in \cref{sec:PSRAndGW}, our ROMs can be performed
at least two orders of magnitude faster than the shooting method.

%%

%---------------------------------------------------------------------
\subsection{Assessing the ROMs of the DEF gravity}
\label{ssec:assessROM}
%---------------------------------------------------------------------

%%
\begin{figure}
  \includegraphics[width=8.66cm]{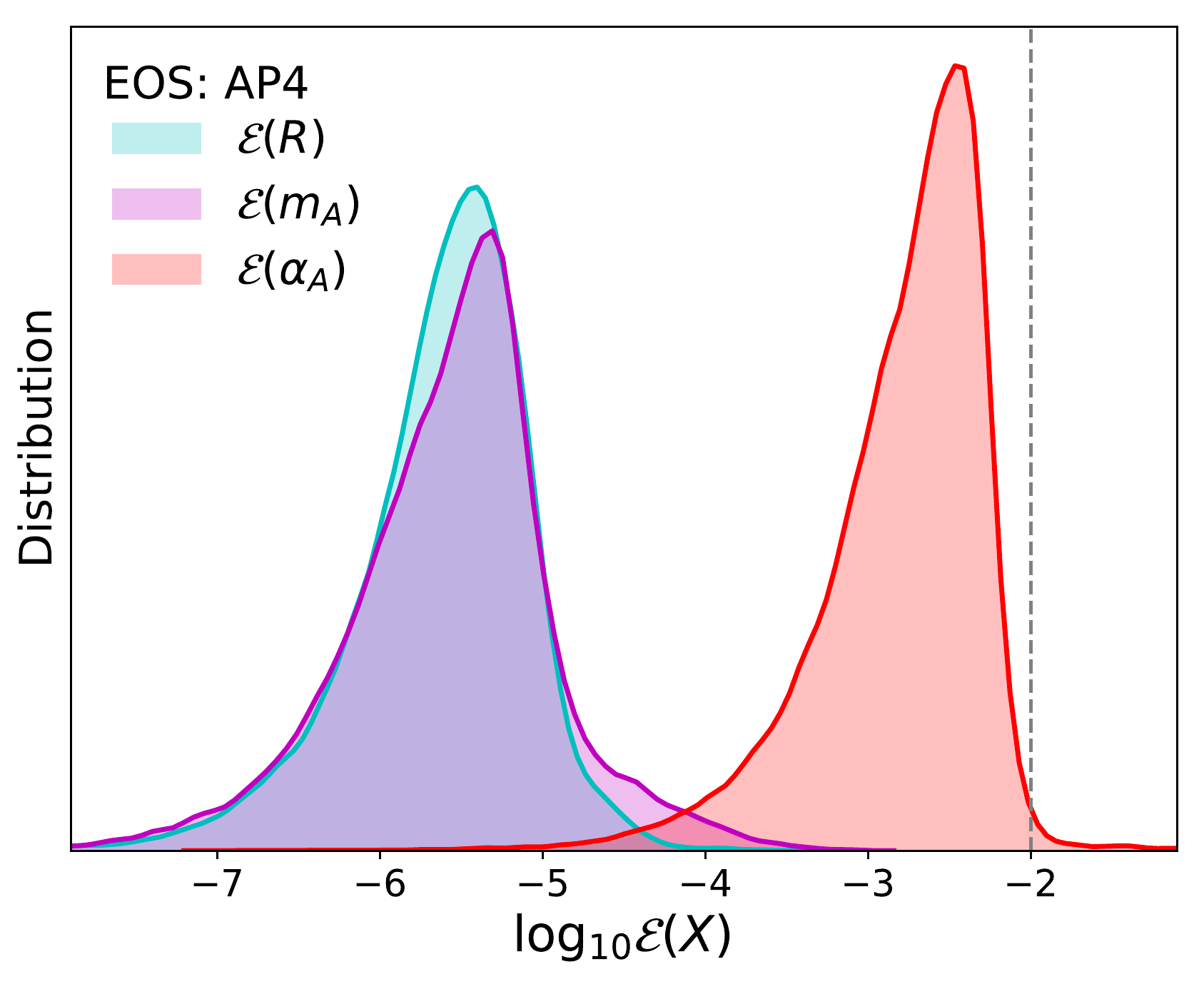}
\caption{(color online) The kernel density estimation (KDE) distributions of the
relative error $\mathcal{E}(X)$, where $X \in \left\{ m_A, R, \alpha_A
\right\}$. The dashed line corresponds to the relative tolerable error
($\lesssim 1\%$) involved in integrating the modified TOV equations in the
shooting method. As the illustration shows, the relative errors of $R$ and $m_A$
are small, $\lesssim 10^{-5}$. In contrast, the upper limit of
$\mathcal{E}(\alpha_A)$ is less than $ 1\%$ as expected, but worse than those of
$R$ and $m_A$. The results are consistent with~\cref{fig:projErr}.
\label{fig:AP4:randrerr}}
\end{figure}

After the ROMs of the DEF gravity are constructed, we need to assess their
accuracy. We define,
\begin{equation}
  \mathcal{E}(X) = \left|\frac{X_{\rm ROM} - X_{\rm mTOV}}{X_{\rm ROM} + X_{\rm mTOV}}
  \right|, 
  \quad X \in \{m_A,\, R,\, \alpha_A\} \,,
  \, \label{eqn:randrerr}
\end{equation}
that measures the fractional accuracy of the ROMs. In~\cref{eqn:randrerr},
$\mathcal{E}(X)$ denotes the relative error with respect to the true value. The
values we predict by the ROMs are denoted as $X_{\rm ROM}$; the values generated
by the shooting algorithm (with a relative tolerable error $\sim 1\%$) are
denoted as $ X_{\rm mTOV}$.

We randomly generate $5\times 10^{5}$ sets of parameters for $\alpha_0$,
$\beta_0$, and $\rho_c$ in the valid ranges of the ROMs. We then generate the
corresponding values for $X_{\rm ROM}$ and $X_{\rm mTOV}$ from the ROMs and the
shooting algorithm respectively. $\mathcal{E}(X)$ is calculated according to
\cref{eqn:randrerr}. The distributions of $\mathcal{E}(X)$ are illustrated
in~\cref{fig:AP4:randrerr}. As the figure shows, the relative errors of $R$ and
$m_A$ are $\lesssim 10^{-5}$. In contrast, $\mathcal{E}(\alpha_A)$ is
approximately three orders of magnitude larger.  The upper limit of
$\mathcal{E}(\alpha_A)$ is less than the tolerable error, $1\%$, of the shooting
method. Therefore, these results are consistent with the relative maximum
projection error discussed in~\cref{fig:projErr}. Notice that, the parameters
used in making \cref{fig:AP4:randrerr} are randomly generated, not necessarily
on the grid in \cref{fig:griddat}. Therefore, the results in
\cref{fig:AP4:randrerr} have included all errors for our ROMs, including the
{\it interpolating} errors.  Although the ROM of $\alpha_A$ leads to a larger
error relative to those of $R$ and $m_A$, the error can be neglected when
compared with the error from the shooting method. It has no noticeable influence
in the parameter estimation that we are to discuss in \cref{sec:PSRAndGW}.

%---------------------------------------------------------------------
\section{Constraints from binary pulsars and gravitational waves}
\label{sec:PSRAndGW}
%---------------------------------------------------------------------

In this section, we apply our ROMs to various scenarios, and discuss the
improvement in deriving NS properties. In these applications, we use the MCMC
technique~\cite{Brooks:2011, ForemanMackey:2012ig, Shao:2017gwu} to constrain
the DEF theory.

\subsection{The setup}

As mentioned before, two independent parameters, $\alpha_0$ and $\beta_0$, are
needed to characterize the DEF theory. One of them, $\alpha_0$, has been well
constrained by the Cassini mission~\cite{Bertotti:2003rm}, that measures the
Shapiro time delay in the weak-field regime. It gives a limit, $|\alpha_0| < 3.4
\times 10^{-3}$ at 68\% confidence level (CL)~\cite{Damour:2007uf}. In the
nonperturbative strong field, the DEF theory can be significantly different from
GR. The constraints in the strong field were studied in
Refs.~\cite{Freire:2012mg, Antoniadis:2013pzd, Wex:2014nva, Shao:2017gwu,
Anderson:2019eay}.

NSs, whose gravitational binding energy can be as large as $\sim20\%$ of their
rest-mass energy, are powerful objects to test the strong-field gravity. We
concentrate on the observations relevant to NSs. We consider (i) systems of
binary pulsars in the quasi-stationary regime, and (ii) GWs from BNSs in the
highly dynamical regime. With the construction of ROMs in~\cref{sec:ROM}, we now
have a numerically faster way to derive NS properties. The current and future
constraints for the DEF theory are investigated in the following.

\begin{table*}
\caption{Binary parameters of the five NS-WD systems (PSRs
J0348+0432~\cite{Antoniadis:2013pzd}, J1012+5307~\cite{Lazaridis:2009kq,
Antoniadis:2016hxz, Desvignes:2016yex}, J1738+0333~\cite{Freire:2012mg},
J1909$-$3744~\cite{Reardon:2015kba, Desvignes:2016yex, Arzoumanian:2017puf} and
J2222$-$0137~\cite{Cognard:2017xyr}) that we use to constrain the DEF theory.
The observed time derivatives of the orbit period, $\dot{P}_b^{\rm obs}$, will
be corrected with the latest Galactic model in Ref.~\cite{2017MNRAS.465...76M}.
The intrinsic derivatives of the orbital period, $\dot{P}_{b}^{\rm int}$, are
obtained from $\dot{P}_b^{\rm obs}$, by subtracting the other kinematic effects,
such as the acceleration effect~\cite{Damour:1990wz} and the ``Shklovskii''
effect~\cite{1970SvA....13..562S}. For PSRs J0348+0432, J1012+5307, and
J1738+0333, the pulsar masses, $m_p^{\rm obs}$, are not derived from ratio
observations. They are obtained from the companion masses, $m_c^{\rm obs}$, and
the mass ratios, $q$, without assuming the validity of GR. Similar consideration 
was made for PSRs~J1909$-$3744 and J2222$-$0137~\cite{Shao:2017gwu}. 
The standard 1-$\sigma$ errors in the least significant
digit(s) are shown in parentheses.}
  \centering
  \def\arraystretch{1.3}
  \begin{tabularx}{\textwidth}{llllll}
    \hline\hline
    Pulsar & J0348+0432~\cite{Antoniadis:2013pzd} &
    J1012+5307~\cite{Lazaridis:2009kq, Antoniadis:2016hxz, Desvignes:2016yex}
    & J1738+0333~\cite{Freire:2012mg} & J1909$-$3744~\cite{Reardon:2015kba,
    Desvignes:2016yex, Arzoumanian:2017puf} &
    J2222$-$0137~\cite{Cognard:2017xyr} \\
    \hline
    Orbital period, $P_b$\,(d) & 0.102424062722(7) & 0.60467271355(3) &
    0.3547907398724(13) & 1.533449474406(13) & 2.44576454(18) \\
    Eccentricity, $e$ & $2.6(9) \times 10^{-6}$ & $1.2(3) \times 10^{-6}$ &
    $3.4(11) \times 10^{-7}$ & $1.14(10) \times 10^{-7}$ & 0.00038096(4) \\
    Observed $\dot P_b$, $\dot P^{\rm obs}_b$\,(${\rm fs\,s}^{-1}$) & $-273(45)$
    & $-50(14)$ & $-17.0(31)$ & $-503(6)$ & $200(90)$ \\
    Intrinsic $\dot P_b$, $\dot P^{\rm int}_b$\,(${\rm fs\,s}^{-1}$) &
    $-274(45)$ & $-5(9)$ & $-27.72(64)$ & $-6(15)$ & $-60(90)$ \\
    Mass ratio, $q\equiv m_p / m_c$ & 11.70(13) & 10.5(5) & 8.1(2) & \ldots &
    \ldots \\
    Pulsar mass, $m_p^{\rm obs}$\,($\SUN$) & \ldots & \ldots & \ldots &
    1.48(3) & 1.76(6) \\
    Companion mass, $m_c^{\rm obs}$\,($\SUN$) & $0.1715^{+0.0045}_{-0.0030}$ &
    0.174(7) & $0.1817^{+0.0073}_{-0.0054}$ & 0.208(2) & $1.293(25)$ \\
    \hline
  \end{tabularx}
  \label{tab:PSR}
\end{table*}

Some stringent limits via the MCMC approach to the spontaneous scalarization of
the DEF theory have been obtained in Ref.~\cite{Shao:2017gwu}. Five asymmetric
NS-WD binaries were used. They are PSRs~J0348+0432~\cite{Antoniadis:2013pzd},
J1012+5307~\cite{Lazaridis:2009kq, Antoniadis:2016hxz, Desvignes:2016yex},
J1738+0333~\cite{Freire:2012mg}, J1909$-$3744~\cite{Reardon:2015kba,
Desvignes:2016yex, Arzoumanian:2017puf} and J2222$-$0137~\cite{Cognard:2017xyr}.
Relevant parameters for these binary pulsars are listed in \cref{tab:PSR}. The
WD companion corresponds to a weakly self-gravitating object. It has a tiny
effective scalar coupling $ \alpha_c \approx \alpha_0$. 
The small effective scalar coupling of the WD would lead to a large dipole
term, $\propto (\alpha_p - \alpha_c)^2$, in the parameter space we are
investigating [see~\cref{eqn:pbdot:dipole}].

Pulsar parameters and orbit parameters are measured by the TOAs of pulses,
including Keplerian and post-Keplerian parameters. Some parameters, such as the
time derivative of the orbital period, $\dot{P}_b$, are functions of the masses
of the pulsar and its companion~\cite{Damour:1991rd}. The chosen systems are
all well timed. The uncertainty in $\dot{P}_b$ can be very small
for the pulsars we consider (see~\cref{tab:PSR}). Such precise measurements
place strong bounds on various alternative theories of
gravity~\cite{Wex:2014nva}.

\begin{table*}
\caption{Properties of GW170817 from the LIGO/Virgo
Collaboration~\cite{TheLIGOScientific:2017qsa, Abbott:2018wiz, Abbott:2018exr,
Abbott:2018lct}. The primary mass $m_1$ and secondary mass $m_2$ are determined
with the low-spin prior assumption~\cite{Abbott:2018wiz}. The radii of NSs,
$R_{1}$ and $R_2$, are derived with the EOS-insensitive
relations~\cite{Yagi:2016bkt,Chatziioannou:2018vzf}.}
    \centering
    \def\arraystretch{1.3}
    \begin{tabularx}{\textwidth}
      {p{3.8cm} p{3.8cm} p{3.8cm} p{3.8cm} p{3.8cm}}
      \hline\hline
        GW170817~\cite{TheLIGOScientific:2017qsa} 
        & $m_1^{\rm obs}\,(\SUN)$~\cite{Abbott:2018wiz} & $m_2^{\rm obs}\,(\SUN)$~\cite{Abbott:2018wiz} 
        & $R_1^{\rm obs}\,({\rm km})$~\cite{Abbott:2018exr} 
        & $R_2^{\rm obs}\,({\rm km})$~\cite{Abbott:2018exr}  \\
      \hline
      90\% CL & $(1.36, 1.60)$ & $(1.16, 1.36)$ & $(9.1, 12.8)$ 
      & $(9.2, 12.8)$ \\
      68\% CL & $(1.41, 1.55)$ & $(1.20, 1.32)$ & $(9.7, 11.9)$  & $(9.6, 11.8)$  \\
      \hline
    \end{tabularx}
    \label{tab:GW}
  \end{table*}

On the other hand, a completely new era for testing highly dynamical strong
field with NSs has began with GW170817~\cite{TheLIGOScientific:2017qsa}. It
offers an extraordinary opportunity, completely different from previously
detected BBH merger events. From the phase of GW170817, one can derive the NS
properties, for instance, the mass and radius of each NS.  The LIGO/Virgo
Collaboration used EOS-insensitive relations to derive those properties in
Ref.~~\cite{Abbott:2018wiz}. The results are shown  in~\cref{tab:GW}, and they
are included to constrain the DEF theory in the MCMC
approach.\footnote{Different from the LIGO/Virgo Collaboration,
\citet{De:2018uhw} used another EOS-insensitive relation. They obtained similar
results.}

In binary pulsars, timing parameters are measured by radio techniques. Some of
them have achieved extremely high precision from decades of observation. In
contrast, the NS properties derived from GW170817, have a relatively poor
precision. But, the radii of NSs were inferred from
GW170817~\cite{TheLIGOScientific:2017qsa,Abbott:2018wiz}. This measurement is
unique to constrain the alternative theories of gravity.

In addition to the systems mentioned above, we expect that, more BNSs are to be
detected with the future GW detectors. KAGRA, operated in Japan, has two 3\,km
orthonormal arms to form an underground GW interferometer. It has a similar
designed sensitivity as that of the Advanced LIGO~\cite{Akutsu:2018axf}.
Moreover, the next-generation ground-based detectors are expected to detect more
GW events with larger SNRs, due to their even higher sensitivities.
CE~\cite{Evans:2016mbw} is an L-shaped 40\,km  interferometer (compared to 4\,km
for LIGO, and 3\,km for Virgo and KAGRA). It is to be built on the experience
and success of current detectors. It is roughly ten times more sensitive than
the Advanced LIGO. Another proposed next-generation GW detector,
ET~\cite{Hild:2010id}, is a 10\,km interferometer. It is supported by the
European Commission. There are three arms forming an equilateral triangle,
located underground to reduce seismic noises. It is also ten times more
sensitive than the Advanced LIGO.  Particularly, in the frequency band below
$\sim10$ Hz, ET can achieve a higher sensitivity than CE~\cite{Evans:2016mbw}.
We want to find out how the DEF theory will be constrained with those future GW
detectors. 

\citet{Shibata:2013pra} discovered the fact that, depending on the EOS one
can still have strong scalarization in a mass range that is not {\it yet}
constrained by pulsar experiments. After accounting for new pulsar tests,
\citet{Shao:2017gwu} showed that there is a so-called ``scalarization window''
at $m_p \sim 1.7\, \SUN$. This is the place where there could still have a large
deviation from GR, given all the current constraints. If the future GW detectors
can observe asymmetric BNSs with masses around $1.6$--$1.7\,\SUN$, this window
could be closed. In this paper, we assume that, a GW event from an inspiraling
BNS with masses, $\left(1.65 \, \SUN, 1.22 \, \SUN \right)$, is detected at
luminosity distance $D_L = 200 \, {\rm Mpc}$. This BNS system has similar NS
masses as those of PSR J1913+1102~\cite{Lazarus:2016hfu}. We denote this
hypothetical BNS as {\sf BNS@200Mpc}. Then, we perform MCMC simulations to
investigate this hypothetical BNS. In the foreseeable future, BNS coalescence
will be observed in large numbers. Correspondingly, a tight constraint on
parameters of the DEF theory, $\alpha_0$ and $\beta_0$, can be obtained.

Some properties of NSs are governed by the EOS. NS EOS describes the relation
between pressure and density of NS matters. It is involved in the modified TOV
equation integration. However, because of our lack of knowledge about the inner
structure of NSs, the EOS is still uncertain. Following our previous discussion,
there exists a lower limit for the maximum mass of NSs, $M_{\rm max} \gtrsim 2
\, \SUN$. Nine EOSs, {\sf AP3, AP4, ENG, H4, MPA1, PAL1, SLy4, WFF1,} and {\sf
WFF2}, are adopted in this study (see Ref.~\cite{Lattimer:2012nd} for a review).
We have illustrated the mass-radius relations of NSs for these EOSs
in~\cref{fig:MRRelation}.  They are all consistent with the above maximum-mass
limit.  Moreover, for our studies, we believe that they are sufficient to
investigate the EOS-dependent properties with spontaneous
scalarization~\cite{Shibata:2013pra}. 

\subsection{The MCMC framework}

Combining the observations of five pulsars, GW170817, and {\sf BNS@200Mpc}, we
perform MCMC simulations with each of these EOSs. MCMC techniques update
posterior distributions of underlying parameters. After convergence, those
distributions will be consistent with astrophysical observations by evaluating
the likelihood function. We use the python implementation of an affine-invariant
MCMC ensemble sampler, {\tt EMCEE}\footnote{\url{https://github.com/dfm/emcee}}.
We use the ROMs, that is built in~\cref{sec:ROM} and coded in {\tt pySTGROM}
package, to speed up the MCMC calculations for parameter estimation within the
Bayesian framework~\cite{DelPozzo:2016ugt,Shao:2017gwu}.

In the Bayesian inference, given priors, the posterior distribution of
$(\alpha_0, \beta_0)$ can be inferred with data, ${\cal D}$, and a hypothesis,
${\cal H}$. We use the formula of Bayes' theorem,
\begin{align}
   P\left(\alpha_0, \beta_0 | {\cal D}, {\cal H}, {\cal I} \right) = 
   \hspace{5.5cm} \nonumber \\
  \int \frac{P\left( {\cal D} | \alpha_0, \beta_0, \bm{\Xi}, 
  {\cal H}, {\cal I}\right) P\left( \alpha_0, \beta_0, \bm{\Xi} | 
  {\cal H},{\cal I}\right)}
  {P\left({\cal D} | {\cal H}, {\cal I} \right)} {\rm d} \bm{\Xi} \,,
  \label{eqn:bayes} 
\end{align}
where $P\left(\alpha_0, \beta_0 | {\cal D}, {\cal H}, {\cal I} \right)$ is an
updated (marginalized) posterior distribution of $(\alpha_0, \beta_0)$, $
P\left( {\cal D} | \alpha_0, \beta_0, \bm{\Xi}, {\cal H}, {\cal I}\right) \equiv
\mathcal{L} $ is the likelihood function, $P\left( \alpha_0, \beta_0, \bm{\Xi} |
{\cal H},{\cal I}\right)$ is the prior on parameters $(\alpha_0, \, \beta_0,\,
\bm{\Xi})$, and $P\left({\cal D} | {\cal H}, {\cal I} \right)$ is the model
evidence. In~\cref{eqn:bayes}, the hypothesis, ${\cal H}$, represents the DEF
theory, ${\cal I}$ denotes all other relevant knowledge, $\bm{\Xi}$ collectively
denotes all other unknown parameters in addition to $(\alpha_0,\, \beta_0)$.

For our studies of the DEF theory, the priors of $(\alpha_0, \beta_0)$ are
chosen carefully to cover the interesting region where the spontaneous
scalarization occurs.  In order to speed up the MCMC simulations, the values of
$(\log_{10} \left|\alpha_0\right|, \beta_0)$ are restricted to the same
rectangle region as in our ROMs that are built in~\cref{ssec:conROM}.  We pick a
uniform prior on $\log_{10} \left|\alpha_0\right|$ in the range $[-5.3,\,-2.5]$.
The prior on $\beta_0$ is chosen to be uniform in the interval $\beta_0 \in
[-4.8,\,-4.0]$. 

The parameters, $\alpha_0$ and $\beta_0$, can be constrained by evaluating the
likelihood function. Due to their observational characteristics, the five
pulsars in \cref{tab:PSR}, GW170817, and {\sf BNS@200Mpc} contribute differently
to the likelihood function.  Their corresponding log-likelihood functions are
given separately as follows.

For binary pulsars, we have the log-likelihood function, 
\begin{align}
\ln \mathcal{L}_{\rm PSR} = 
\hspace{7.cm} \nonumber \\
- \frac{1}{2} \sum_{i=1}^{N_{\rm PSR}} 
\left[ \left(\frac{ \dot P_b^{\rm th} - 
\dot P_b^{\rm int}}{\sigma_{\dot P_b^{\rm int}}} \right)^2 +
\left( \frac{m_{p} - m_{p}^{\rm obs}}{\sigma_{m_p^{\rm obs}}} \right)^2 + 
\left( \frac{m_{c} - m_{c}^{\rm obs}}{\sigma_{m_c^{\rm obs}}} \right)^2 \right]_i 
  \,, \label{eqn:loglike:Pulsar} 
\end{align}
for $N_{\rm PSR}$ binary pulsar systems (we have $N_{\rm PSR}=5$ in this study).
We have assumed that observations with different binary pulsars are independent.
For each pulsar, the intrinsic orbital decay, $\dot{P}_b^{\rm int}$, the pulsar
mass, $m_p^{\rm obs}$, and the companion mass, $m_c^{\rm obs}$ are given
in~\cref{tab:PSR} when applicable.  In some cases, we have the mass ratio $q
\equiv m_p/m_c$ and $m_c^{\rm obs}$ instead; it is easy to obtain $m_p^{\rm
obs}$.  The 1-$\sigma$ uncertainty, $\sigma_X$ in \cref{eqn:loglike:Pulsar}, is
the observational uncertainty for the quantity $X$, where $X \in \left\{\dot
P_b^{\rm int}, m_p^{\rm obs}, m_c^{\rm obs} \right\}$. The predicted orbital
decay, $\dot P_b^{\rm th}$, from the DEF theory equals to $\dot P_b^{\rm dipole}
+ \dot P_b^{\rm quad}$ [see~\cref{eqn:pbdot:dipole,eqn:pbdot:quad}]. The
quantities, $\dot P_b^{\rm th}$ and $m_p$, are implicitly dependent on the
parameter set, $\left(\alpha_0, \beta_0, \bm{\Xi}\right)$. During the MCMC
simulations, they are derived from those parameters. The companion mass, $m_c$,
is picked randomly within its 1-$\sigma$ uncertainty.

For the full calculation, it is worth noting that, the orbital period, $P_b$,
and the orbital eccentricity, $e$, should also be included
in~\cref{eqn:loglike:Pulsar}. Those quantities have been determined very well
from the observations. We adopt their central values directly for simplifying
the MCMC processes. It is verified that, the uncertainties of them have little
effect on our final limits of $(\alpha_0, \beta_0)$. 

In general, the log-likelihood function for $N_{\rm GW}$ GWs can be expressed as,
\begin{align}
  \ln \mathcal{L}_{\rm GW} & = - \frac{1}{2} \sum_{i=1}^{N_{\rm GW}} \sum_{j=1,2}\left[
    \left( \frac{m_{j} - m_{j}^{\rm obs}}{\sigma_{m_{j}^{\rm obs}}} \right)^2
    +
    \left( \frac{R_{j} - R_{j}^{\rm obs}}{\sigma_{R_{j}^{\rm obs}}} \right)^2 \right]_i
    \nonumber \\
    & \quad + \sum_{i=1}^{N_{\rm GW}} {\tt D}_i 
    \left(|\alpha_1 - \alpha_2|, \, |\Delta \alpha|^{\rm upper} \right) \,. \label{eqn:loglike:GW}
\end{align}
Here, the properties of BNSs, such as the masses, $m_j$ $(j=1,2)$, the radii,
$R_j$, are given in~\cref{tab:GW}. The 1-$\sigma$ uncertainties of those
properties, $\sigma_{m_j^{\rm obs}}$ and $\sigma_{R_j^{\rm obs}}$, are given at
the 68\% CL. The function for the dipole radiation, ${\tt
D}_i \left(|\alpha_1 - \alpha_2|, \, |\Delta \alpha|^{\rm upper} \right)$ ,
describes the contribution of the $i$-th GW's dipolar radiation to $\ln \cal
L_{\rm GW}$.  It returns 0, when the effective scalar couplings of the BNS,
$\alpha_1$ and $\alpha_2$, satisfy $\left|\alpha_1 - \alpha_2\right| =
\left|\Delta \alpha\right| \leq \left|\Delta \alpha\right|^{\rm upper}$,
otherwise it returns $-\infty$. Here $\left|\Delta \alpha\right|^{\rm upper}$ is
the upper limit of the absolute difference between $\alpha_1$ and $\alpha_2$
from observations. It is derived from the relation, $B = 5 \left| \Delta \alpha
\right|^2 / 96$~\cite{Barausse:2016eii}, where $B$ is a theory-dependent
parameter regulating the strength of the dipolar radiation in scalar-tensor
theories. The presence of dipole radiation in GW170817 is constrained to be $B
\leq 1.2 \times 10^{-5}$~\cite{Abbott:2018lct}. Therefore, the upper bound of
$\left|\Delta \alpha \right|$, $\left|\Delta \alpha \right|^{\rm upper} \simeq
0.015$, is obtained for GW170817. Notice that we are using NS
masses and radii derived from GR. Because here we are {\it constraining}
the non-GR effects, we feel our approach sufficient. However, at the stage
of {\it discovering} non-GR effects, we need a fully consistent approach
that uses NS masses and radii derived from the scalar-tensor gravity,
instead of GR.

In addition to GW170817, we introduce a hypothetical GW from an inspiraling
BNS, {\sf BNS@200Mpc}, for forecasting future constraints. It is meaningful to
investigate the improvement with future GW detectors. The log-likelihood
function of $N_{\rm GW}^{\star}$ hypothetical GWs is chosen as,
\begin{equation}
  \ln \mathcal{L}_{\rm BNS} = - \frac{1}{2} \sum_{i=1}^{N_{\rm GW}^{\star}} 
  \left[ \left( \frac{\alpha_1 - \alpha_2 }{ \sigma (|\Delta \alpha|)} \right)^2 \right]_i
    \,. \label{eqn:loglike:BNS}
\end{equation}
In~\cref{eqn:loglike:BNS}, $\sigma(|\Delta \alpha|)$ is the expected 1-$\sigma$
uncertainty of $\left| \Delta \alpha \right|$. It is obtained from the approach
of the Fisher information matrix. We will introduce this approach briefly and
perform the MCMC calculations with the log-likelihood function, $\ln
\mathcal{L}_{\rm BNS}$, in~\cref{ssec:future_measure}.  Compared with
\cref{eqn:loglike:GW}, we neglect the contributions from masses and radii,
because, given the starting frequency of CE and ET, the dipolar radiation
contribution is more constraining.

In a short summary, the likelihood function we use in~\cref{eqn:bayes} is the sum
of~\cref{eqn:loglike:Pulsar,eqn:loglike:BNS,eqn:loglike:GW} that have included
all contributions from binary pulsars, GW170817, and future BNSs.

Now, we will explain how to employ the MCMC technique to get the posteriors from
the priors on $(\alpha_0, \beta_0)$ and the log-likelihood functions. Combining
the observations of binary pulsars and GWs, we have $N=N_{\rm PSR} + 2\,
\left(N_{\rm GW} + N_{\rm GW}^{\star} \right)$ NSs to constrain the $(\alpha_0,
\beta_0)$ parameter space. To fully describe the contribution of the
gravitational dipolar radiation of these systems, $N + 2$ free parameters,
denoted collectively as $\bm{\theta}$, are required. The parameters,
$\bm{\theta}$, include $\alpha_0$, $\beta_0$, and $\rho_c^{(i)} (i = 1, \ldots,
N)$~\cite{Shao:2017gwu}. The initial central matter densities, $\rho_c^{(i)}$,
are needed in the Jordan frame. They are fed to our ROMs to derive the NS
properties.  Initially they will be sampled around their GR values, but they are
allowed to explore a sufficiently large range in the MCMC process. In principle,
if we were using the shooting method, the initial central values of the scalar
field, $\varphi_c^{(i)}$, are needed as well. In our ROMs, they are no longer
involved. Those NS properties can be obtained uniquely with $\alpha_0$,
$\beta_0$, and $\rho_c^{(i)}$.  Then, those derived properties are passed to the
log-likelihood
functions,~\cref{eqn:loglike:Pulsar,eqn:loglike:BNS,eqn:loglike:GW}, to evaluate
the posteriors.

The constraints from the same five binary pulsars have been investigated in
great detail in Ref.~\cite{Shao:2017gwu}. In this paper, in order to verify the
validity of our newly built ROMs, we will reproduce their result. In addition to
the five pulsars, we also use GW170817 and a hypothetical BNS, {\sf
BNS@200Mpc}. The {\sf BNS@200Mpc} is assumed to be detected by the Advanced LIGO
at its designed sensitivity, and next-generation detectors, CE and ET. Now, we
can use those observations to constrain the DEF theory, and obtain the bounds in
the $(\alpha_0, \beta_0)$ parameter space. Furthermore, we also quantify
how much the tests will be improved with those next-generation detectors.

\begin{table*}
\caption{The different scenarios used in the paper to constrain the DEF theory.
Different combinations of five binary pulsars (see~\cref{tab:PSR}), GW170817
(see~\cref{tab:GW}), and a hypothetical BNS {\sf BNS@200Mpc}, are investigated.
Their corresponding log-likelihood functions, $\ln \mathcal{L}_{\rm PSR}$, $\ln
\mathcal{L}_{\rm GW}$, and $\ln \mathcal{L}_{\rm BNS}$, are expressed
in~\cref{eqn:loglike:BNS,eqn:loglike:GW,eqn:loglike:Pulsar}.  The scenarios (i)
to (iii) correspond to real observations, while the scenarios (iv) to (vi)
involve a hypothetical {\sf BNS@200Mpc} to be observed by the Advanced LIGO, CE,
and ET. Notice that the scenario (ii) was investigated in great detail in 
Ref.~\cite{Shao:2017gwu}, while the other five scenarios include new extensions. 
\label{tab:scenarios}}
      \centering
      \def\arraystretch{1.3}
      \begin{tabularx}{\textwidth}
        {p{0.9cm} p{3cm} p{3.5cm} p{12cm}}
        \hline\hline
        & Scenario & Log-likelihood function & \\
        \hline
        (i) & {\sf GW170817} & $\ln \mathcal{L}_{\rm GW}$ & GW170817 only\\
        (ii) & {\sf PSRs} & $\ln \mathcal{L}_{\rm PSR}$ & five pulsars \\
        (iii) & {\sf PSRs+GW170817} & $\ln \mathcal{L}_{\rm PSR} + \ln \mathcal{L}_{\rm GW}$ &
combining five pulsars and GW170817 \\ 
        (iv) & {\sf PSRs+aLIGO} & $\ln \mathcal{L}_{\rm PSR} + \ln \mathcal{L}_{\rm BNS,aLIGO}$ &
combining five pulsars and a {\sf BNS@200Mpc} observed by the Advanced LIGO \\
        (v) & {\sf PSRs+CE} & $\ln \mathcal{L}_{\rm PSR} + \ln \mathcal{L}_{\rm BNS,CE}$ &
combining five pulsars and a {\sf BNS@200Mpc} observed by CE \\ 
        (vi)& {\sf PSRs+ET} & $\ln \mathcal{L}_{\rm PSR} + \ln \mathcal{L}_{\rm BNS,ET}$ &
combining five pulsars and a {\sf BNS@200Mpc} observed by ET \\ 
        \hline
      \end{tabularx}
\end{table*}

For each EOS, we perform six separate MCMC runs with different combinations of
observations. These six scenarios are shown in~\cref{tab:scenarios}. Here the
investigations of the DEF theory are divided into two catalogs.
\begin{itemize}
    \item Scenarios (i) to (iii) correspond to the present observations that
    contain five pulsars and GW170817. We will compare the constraining
    results of five pulsars with those of GW170817. The corresponding
    constraints from binary pulsars and GW170817 are given
    in~\cref{ssec:PSRAndGW170817}.
    \item Scenarios (iv) to (vi) involve five pulsars and a hypothetical {\sf
    BNS@200Mpc} to be observed by different GW detectors. We expect to obtain
    tighter constraints compared with present observations. Corresponding
    constraints from binary pulsars and a hypothetical BNS are investigated
    in~\cref{ssec:future_measure}.
\end{itemize}

In the following subsections, we use the EOS {\sf AP4} as an example. At the end
of this section, a total of 54 MCMC runs (6 scenarios $\times$ 9 EOSs) are
discussed.

For each run, we produce 2.6 millions of samples in total using multiple chains
(26 chains $\times \, 10^{5}$ samples for each chain). According to the guides
in Refs.~\cite{Brooks:2011,ForemanMackey:2012ig}, the first half chain samples
of these 54 runs are discarded as the {\sc burn-in} phase.  Then, we ``thin''
remaining samples, with a thinning factor of ten, to reduce the correlation of
adjacent points. Finally, there are $1.3 \times 10^{5}$ ``thinned'' samples
remaining for each scenario. In our studies, it is verified that, those
``thinned'' samples have passed the Gelman–Rubin convergence diagnostic very
well~\cite{Gelman:1992zz}. It indicates that, all parameters in $\bm{\theta}$
have lost the memory of their initial values, such that they can be used to
infer the parameters including $(\alpha_0, \beta_0)$.

%---------------------------------------------------------------------
\subsection{Constraints from binary pulsars and GW170817}
\label{ssec:PSRAndGW170817}
%---------------------------------------------------------------------

In this subsection, we investigate three scenarios with real observations, {\sf
GW170817}, {\sf PSRs}, and {\sf PSRs+GW170817}. We use the EOS {\sf AP4} as an
example. Following our previous discussion, we distribute initial values of
$\log_{10}|\alpha_0|$ and $-\beta_0$ uniformly in the rectangle region of the
parameter space. Then, after the MCMC simulations, we obtain the posterior
distributions, from where we can get upper limits of theory parameters,
$\alpha_0$ and $\beta_0$.

\begin{figure}
  \includegraphics[width=8.66cm]{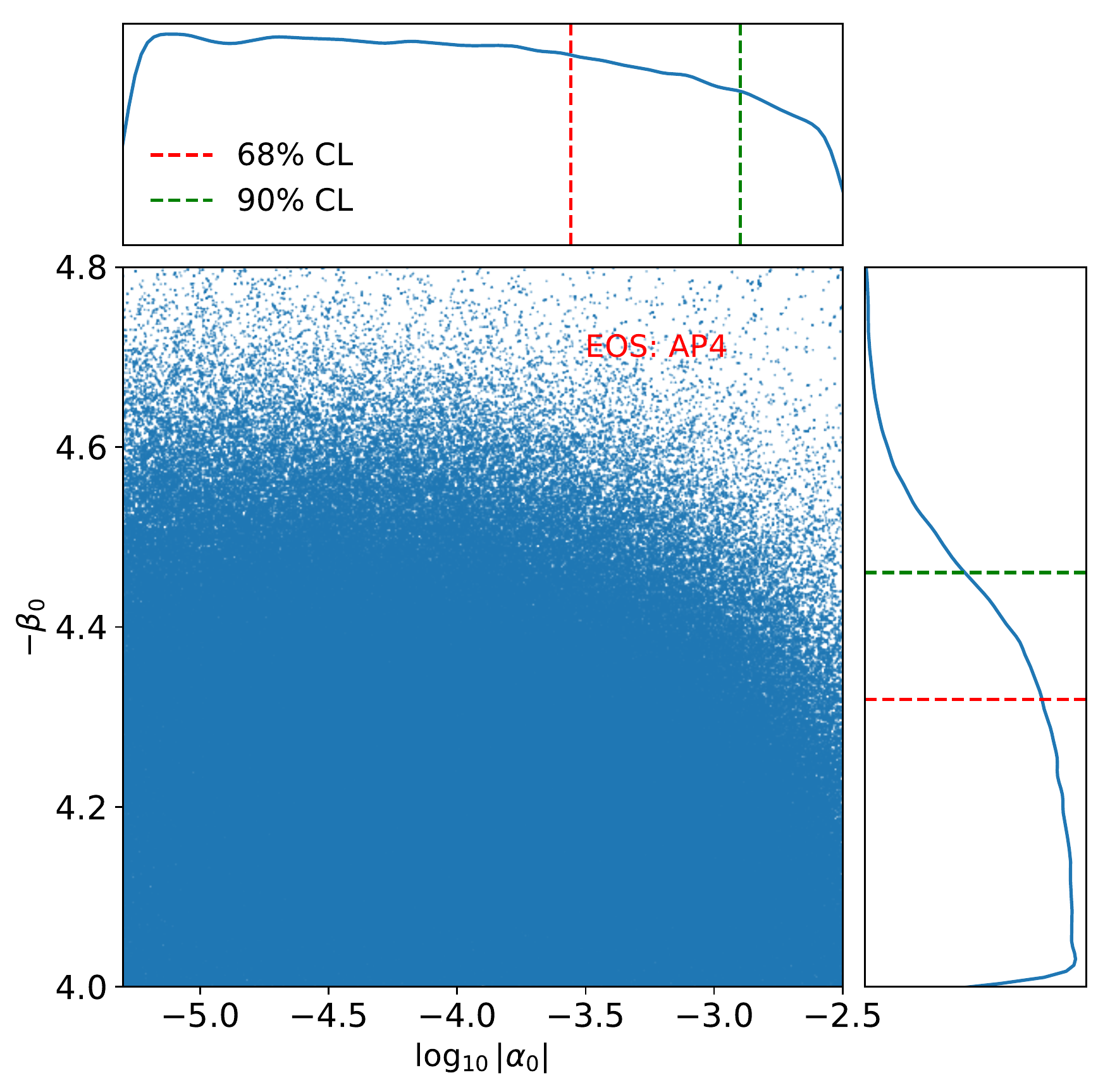}
\caption{(color online) The marginalized 2-dimensional distribution in the
parameter space of $(\log_{10}|\alpha_0|, -\beta_0)$ in the scenario {\sf
GW170817}, for the EOS {\sf AP4}. The marginalized 1-dimensional KDE
distribution of $\log_{10} |\alpha_0|$ is illustrated in the upper panel, while
that of $- \beta_0$ is illustrated in the right panel. The upper limits of
parameters are shown in the red dashed lines at 68\% CL and in the green dashed
lines at 90\% CL. The ``sharp dives'' in the marginalized 1-dimensional 
distributions near the boundary are the ``boundary effects'' of the KDE process; 
same for the other figures that use KDE.
\label{fig:AP4:GW}}
\end{figure}

First, we investigate the scenario {\sf GW170817}. The properties of GW170817
are given in~\cref{tab:GW}. They are used to perform the MCMC simulations.  The
marginalized 2-dimensional distribution in the parameter space of
$(\log_{10}|\alpha_0|, -\beta_0)$ is illustrated in~\cref{fig:AP4:GW}. It is
evident that, GW170817 almost has no constraint on $\alpha_0$. Actually, it is
within our expectation. Because of its short duration, compared with binary
pulsar observations, the parameters are derived with lower precision. Current
data are not accurate enough to give tight constraints. Nevertheless, the upper
limit of $-\beta_0$ is constrained to be $\lesssim 4.5$ at 90\% CL. It is
consistent with the argument that, $\beta_0$ plays the major role in controlling
the strength of scalarization. Therefore, $\beta_0$ is constrained.

\begin{figure}
  \includegraphics[width=8.66cm]{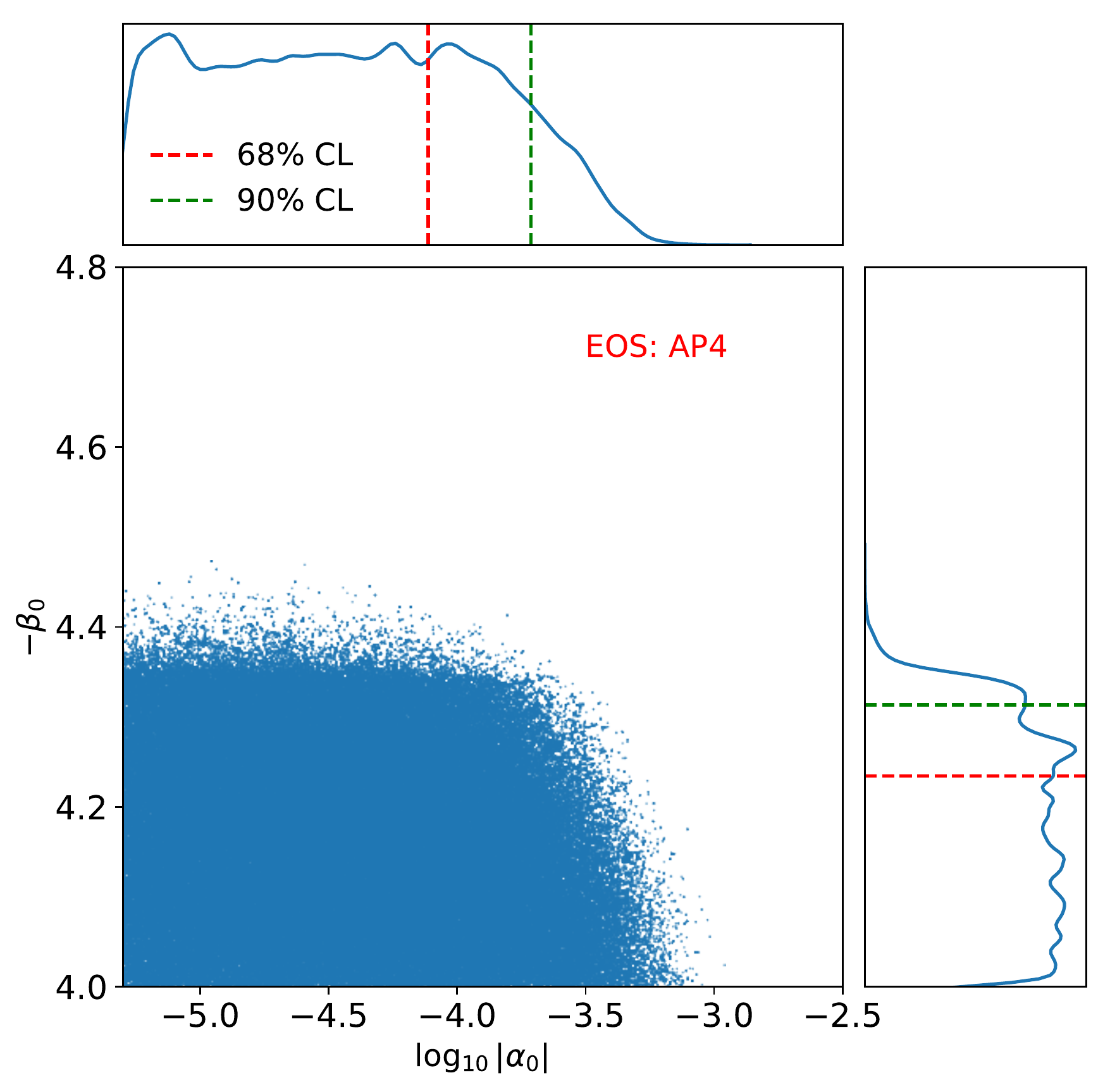}
\caption{(color online) Same as \cref{fig:AP4:GW}, for the scenario {\sf
PSRs}. \label{fig:AP4:PSR}}
\end{figure}

We use five binary pulsars for the scenario {\sf PSRs}. Different from the
illustration in~\cref{fig:AP4:GW}, the results of the scenario {\sf PSRs} show
very good constraints on the parameters of the DEF theory in~\cref{fig:AP4:PSR}.
They are consistent with results in Ref.~\cite{Shao:2017gwu} (in particular, we 
compared our results with Table II of Ref.~\cite{Shao:2017gwu}).
In~\cref{fig:AP4:PSR}, the posterior samples are gathered in the lower left
corner in the marginalized 2-dimensional distribution. The corresponding
parameters, $\log_{10}|\alpha_0|$ and $-\beta_0$, are constrained. Especially,
the parameter $\alpha_0$ is constrained tightly in this scenario within our MCMC
setting. The upper limit of it achieves the level of $\lesssim 10^{-3.7}$ at
90\% CL. Compared with the constraint in~\cref{fig:AP4:GW}, it dictates that,
binary pulsar observations can lead to better constraints on $\alpha_0$ than
GW170817. Moreover, the bound on the parameter $\beta_0$ becomes tighter as
well. The right and upper parts of the parameter space have no support from MCMC
samples.

It is worth noting that, the non-smoothness of the 1-dimensional marginalized
distributions, e.g. in \cref{fig:AP4:PSR}, is caused by the statistical
fluctuations in the MCMC simulations. It has no statistical bearing, and will
disappear if we have infinite samples. As we noted before, our samples have
passed the Gelman–Rubin test for convergence, thus our limits on $\alpha_0$ and
$\beta_0$ are reliable.

\begin{figure}
  \includegraphics[width=8.66cm]{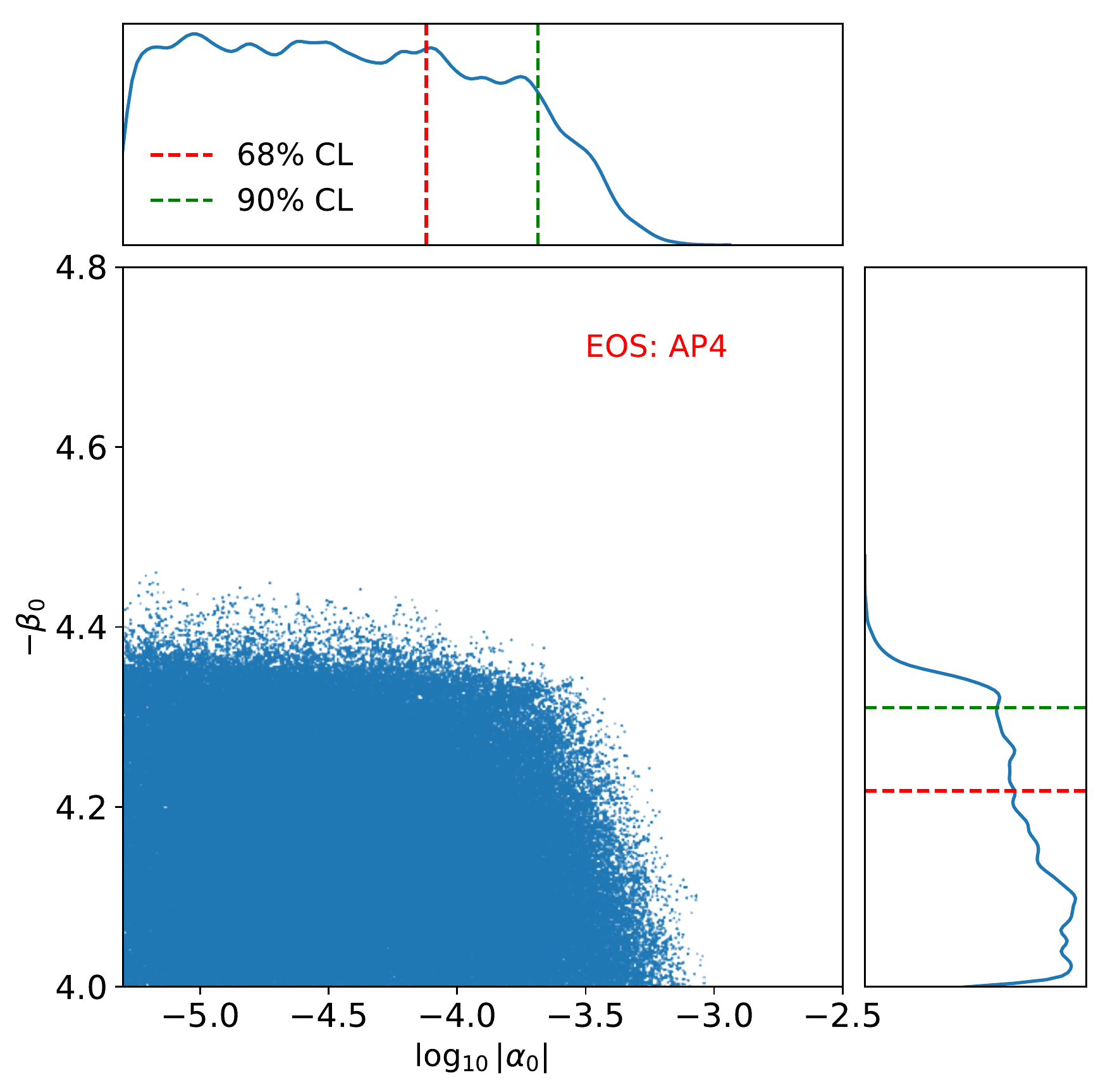}
\caption{(color online) Same as \cref{fig:AP4:GW}, for the scenario {\sf
PSRs+GW170817}. \label{fig:AP4:PSRGW} }
\end{figure}

In~\cref{fig:AP4:PSRGW}, we illustrate the results for the scenario {\sf
PSRs+GW170817}. It involves the combination of the observations of five pulsars
and GW170817. Basically, its result is consistent with the conclusion
in~\cref{fig:AP4:PSR}. It is verified again, that the observation of GW170817
has little effect in constraining the DEF theory. In contrast to GW170817, the
contemporary observation of radio pulsars gives us a quite powerful tool in
probing the strong-field regime for gravity.

%---------------------------------------------------------------------
\subsection{Constraints from binary pulsars and a hypothetical BNS}
\label{ssec:future_measure} 
%---------------------------------------------------------------------

In the previous subsection, we find that, compared with binary pulsars, GW170817
has little effect in constraining the DEF theory.  It is interesting to
investigate whether future ground-based GW detectors can surpass binary pulsar
observations. We investigate the other three scenarios in \cref{tab:scenarios},
namely, {\sf PSRs+aLIGO}, {\sf PSRs+CE}, and {\sf PSRs+ET}. We assume that, the
hypothetical event, {\sf BNS@200Mpc}, has component masses $(1.65\,\SUN$,
$1.22\,\SUN)$. One of the masses is chosen to be close to the ``scalarization
window''~\cite{Shao:2017gwu} where, given the binary pulsar observations, there
could still exist a large deviation from GR.

\begin{table}
\caption{The first column lists three ground-based GW detectors. The second
column gives their frequencies at low end. The third and fourth columns are
respectively the expected SNRs and the expected 1-$\sigma$ uncertainties in the
difference of the effective scalar couplings from the Fisher-matrix
analysis~\cite{Shao:2017gwu}, for a hypothetical event, {\sf
BNS@200Mpc}.\label{tab:Delta}}
    \centering
    \def\arraystretch{1.2}
    \setlength{\tabcolsep}{0.5cm}
    \begin{tabular}{lllll}
  \hline\hline
  Detector & $f_{\rm in}$ (Hz) & $\rho$ & $ \sigma \left(\left| \Delta \alpha
  \right|\right)$ \\ \hline
  aLIGO & 10 & 10.6 & $8.8 \times 10^{-3}$ \\
  CE & 5 & 450 & $7.9 \times 10^{-4}$ \\
  ET & 1 & 153 & $4.0 \times 10^{-4}$ \\
    \hline
  \end{tabular}
  \end{table}

We investigate the power spectral density (PSD) $S_n(f)$ in GW
detectors~\cite{Evans:2016mbw}. The SNR of a GW event is $\rho = \left( \tilde
h(f) \right| \left. \tilde h(f) \right)^{1/2}$, where $\tilde{h}(f)$ is a
Fourier-domain waveform, and the inner product is defined as~\cite{Finn:1992wt},
%--
\begin{equation}
  \label{innerprod}
  \left(\tilde{h}_{1}(f) \right| \left. \tilde{h}_{2}(f)\right) 
  \equiv 2 \int_{f_{\min }}^{f_{\max }} \frac{\tilde{h}_{1}^{*}(f) 
  \tilde{h}_{2}(f)+\tilde{h}_{1}(f) \tilde{h}_{2}^{*}(f)}{S_{n}(f)} \mathrm{d} f \,.
\end{equation} 
%--
Here, the initial frequency $f_{\min}$ is chosen as the starting frequency,
$f_{\rm in}$, for a GW detector, the final frequency is set to be twice of the
frequency of the innermost stable circular orbit (see Ref.~\cite{Shao:2017gwu}).
Those starting frequencies and SNRs are listed in~\cref{tab:Delta}.

In~\cref{eqn:loglike:BNS}, the expected 1-$\sigma$ uncertainties of $\left|
\Delta \alpha \right|$, denoted as $\sigma \left( \left| \Delta \alpha \right|
\right)$, are needed for different GW detectors. The technique of Fisher
information matrix is used to calculate them. The Fisher information matrix is a
measure of an experiment's resolving power for the waveform parameters,
collectively denoted as $\bm{\xi}$. It is defined as~\cite{Finn:1992wt},
%--
\begin{equation}
  \mathcal{I}_{i j}=\left(\frac{\partial \tilde{h}}{\partial \xi_{i}} 
  \left| \frac{\partial \tilde{h}}{\partial \xi_{j}} \right. \right)
  \label{eqn:fisher} \,.
\end{equation}
%--
See Appendix A in Ref.~\cite{Shao:2017gwu} for more details.

For a parameter $\xi_i$, a lower bound on its variance expected from an
experiment can be placed with the inequality of the Cram{\'e}r-Rao
bound~\cite{Cramer:1946,Rao:1992}, $\sigma\left(\xi_{i}\right) \geq
\sqrt{\left(\mathcal{I}^{-1}\right)_{i i}}$. A lower bound on $\left| \Delta
\alpha \right|$ can be got from the diagonal component of the inverse Fisher
matrix.  The corresponding 1-$\sigma$ uncertainties of $\left| \Delta \alpha
\right|$ for the GW detectors are given in~\cref{tab:Delta}. The limits from the
next-generation detectors will be better than that of the Advanced LIGO, due to
better low-frequency sensitivities.

\begin{figure}
  \includegraphics[width=8.66cm]{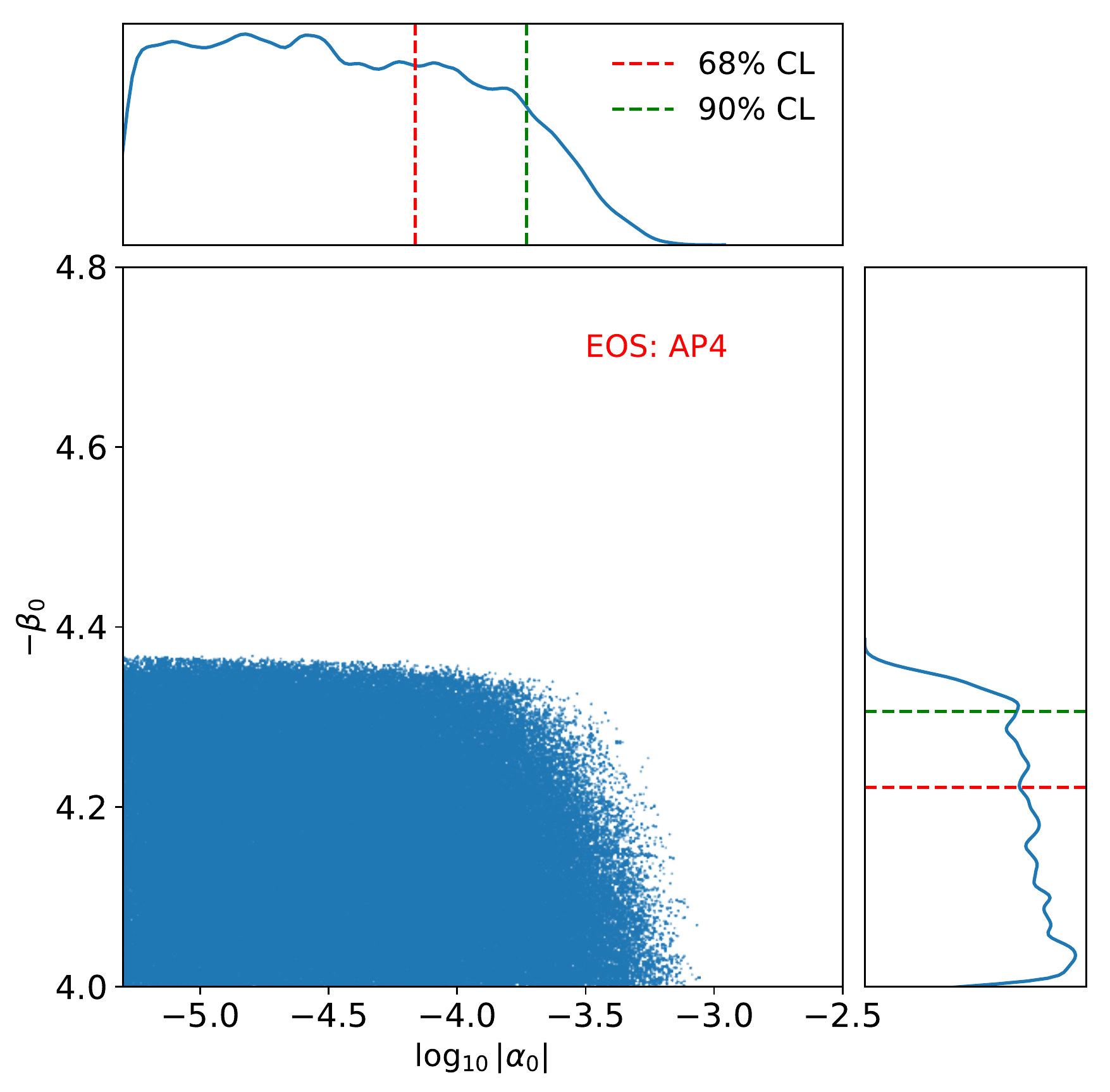}
\caption{(color online) Same as \cref{fig:AP4:GW}, for the scenario {\sf
PSRs+aLIGO}.\label{fig:AP4:PSRaLIGO}}
\end{figure}
\begin{figure}
  \includegraphics[width=8.66cm]{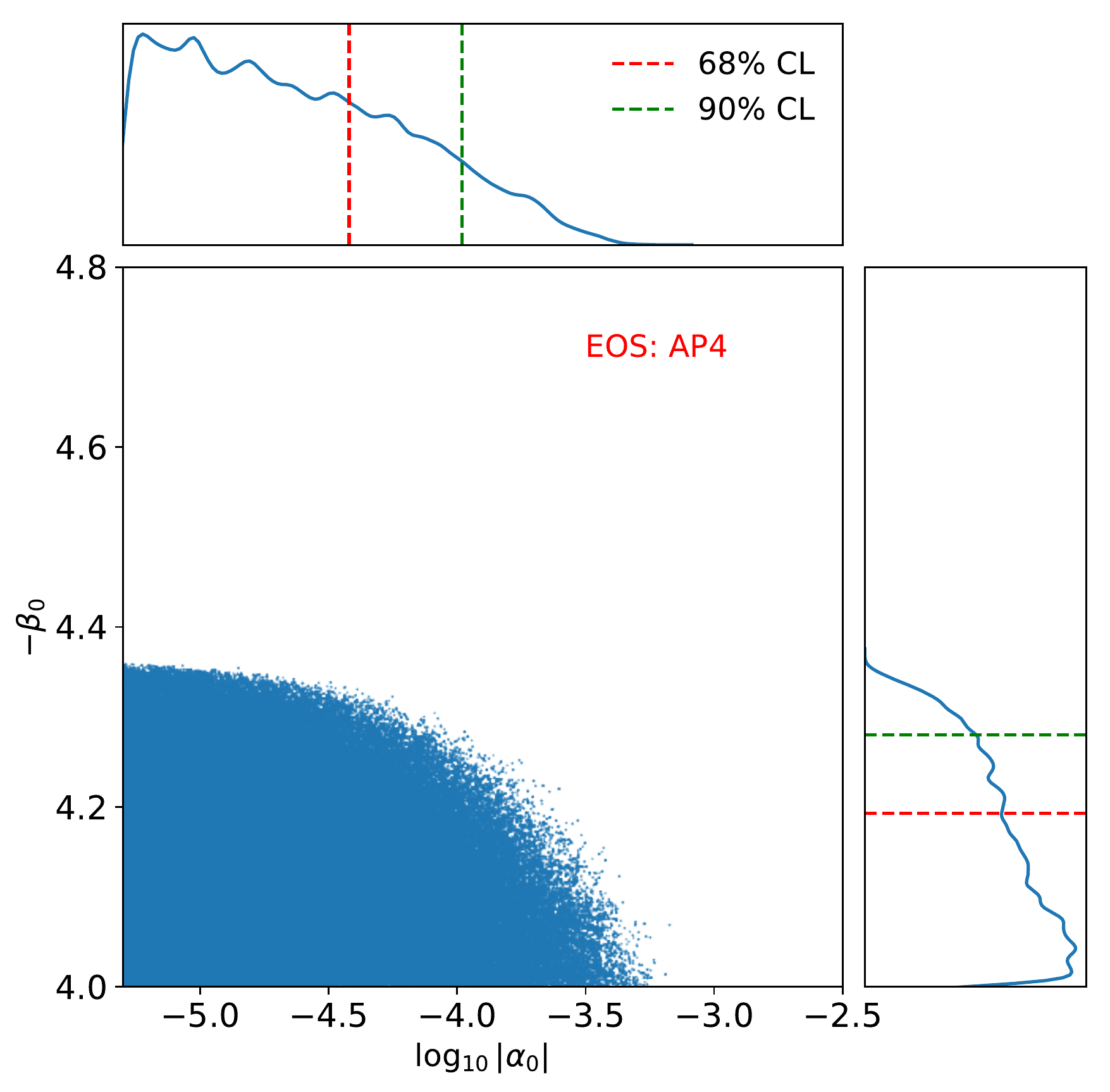}
\caption{(color online) Same as \cref{fig:AP4:GW}, for the scenario {\sf
PSRs+CE}.\label{fig:AP4:PSRCE}}
\end{figure}
\begin{figure}
  \includegraphics[width=8.66cm]{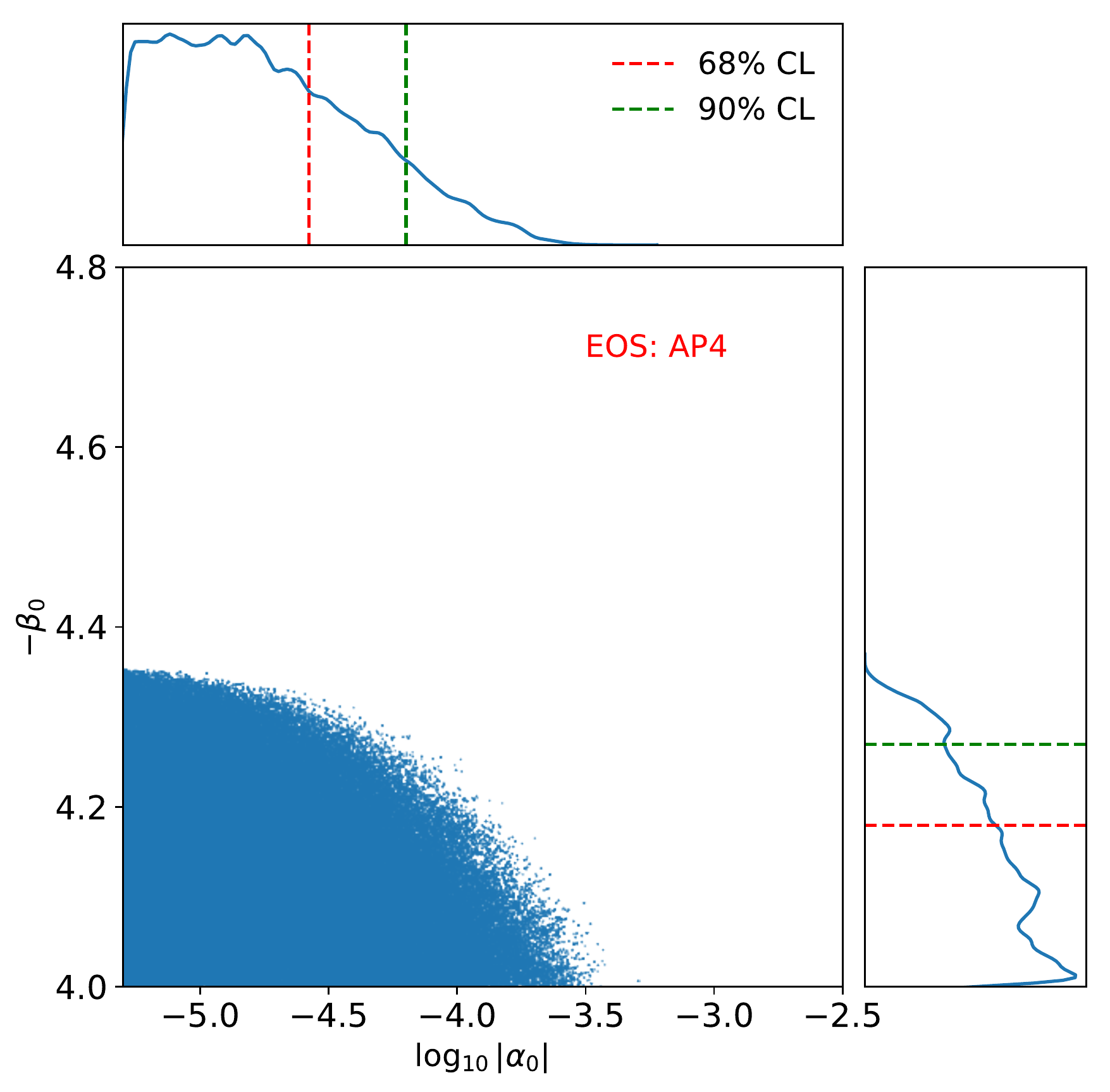}
\caption{(color online) Same as \cref{fig:AP4:GW}, for the scenario {\sf
PSRs+ET}. \label{fig:AP4:PSRET}}
\end{figure}

In the following, we give the results for the scenarios {\sf PSRs+aLIGO}, {\sf
PSRs+CE}, and {\sf PSRs+ET}.

For the scenario {\sf PSRs+aLIGO}, in addition to the five binary pulsars, we
use a hypothetical {\sf BNS@200Mpc}, assumed to be observed by the Advanced
LIGO. We illustrate the result of this scenario in~\cref{fig:AP4:PSRaLIGO}.
Compared with the scenario {\sf PSRs+GW170817} in~\cref{fig:AP4:PSRGW}, it is
evident that the constraints on the DEF theory are not significantly improved.

The scenarios, {\sf PSRs+CE} and {\sf PSRs+ET} for the EOS {\sf AP4}, are
illustrated respectively in~\cref{fig:AP4:PSRCE} and~\cref{fig:AP4:PSRET}. The
parameter space of the DEF theory is highly constrained with those
next-generation detectors. We can obtain the tightest constraints on the
parameters, $\alpha_0$ and $\beta_0$, in our studies. Here, the upper limit of
$\left| \alpha_0 \right|$ achieves the level of $\lesssim 10^{-4.0}$ at 90\% CL,
given our priors. The parameter, $-\beta_0$, can be constrained to $\lesssim
4.3$ at 90\% CL. Due to its low-frequency sensitivity, ET can track a much
longer time of a BNS inspiral than the Advanced LIGO and CE. It will constrain
the DEF theory better.

\begin{figure}
  \centering
  \includegraphics[width=8.66cm]{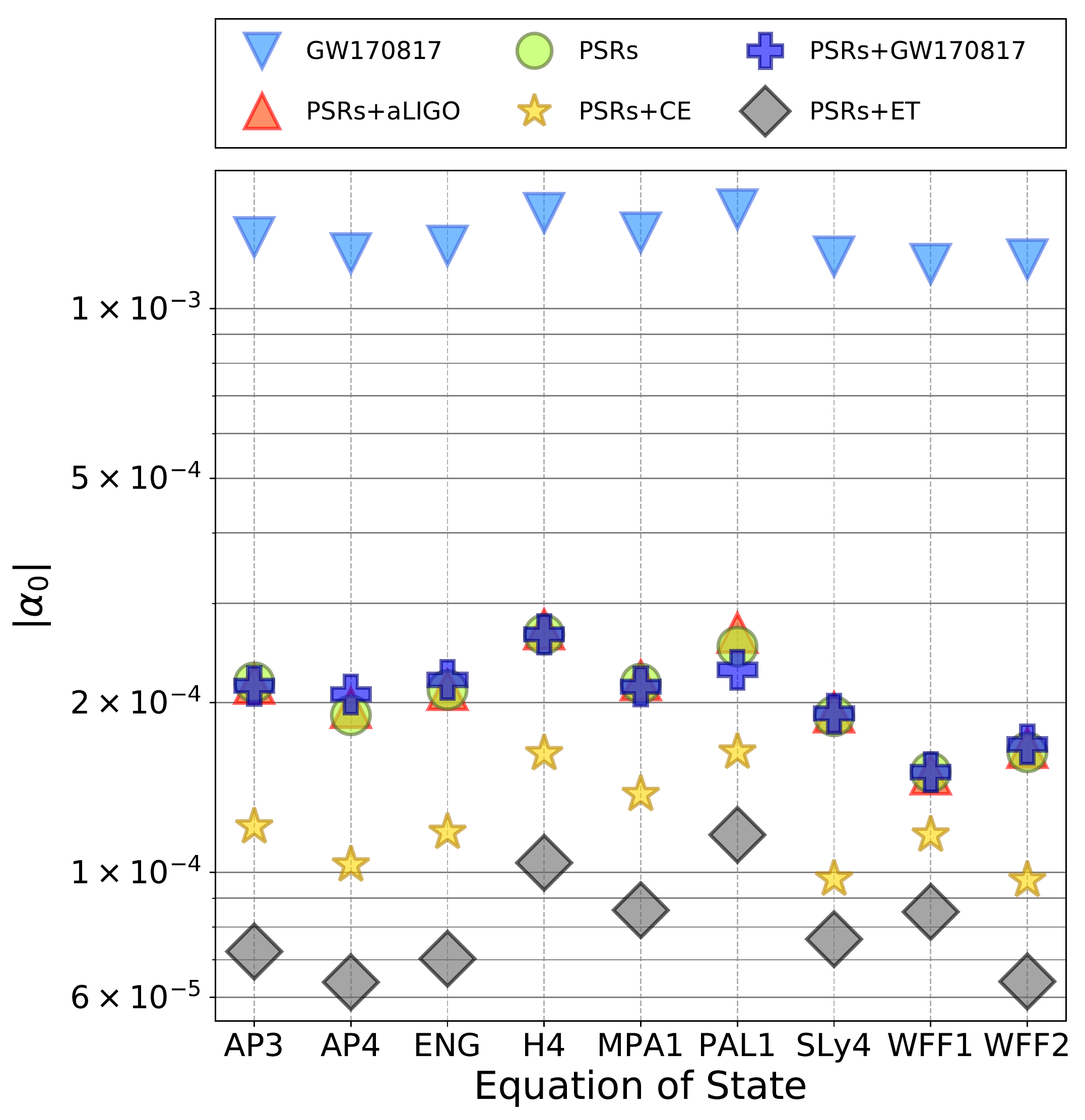}
\caption{(color online) 90\% CL upper bounds on the parameter, $\left|
\alpha_0 \right|$, for nine EOSs and six scenarios in our studies. Their
marginalized 1-dimensional distributions are shown
in~\cref{fig:app:GW,fig:app:Fisher}. Notice that the limit on $\left| \alpha_0
\right|$ is influenced by our priors (see text).
\label{fig:alpha_90CL}}
\end{figure}
\begin{figure}
  \centering
  \includegraphics[width=8.66cm]{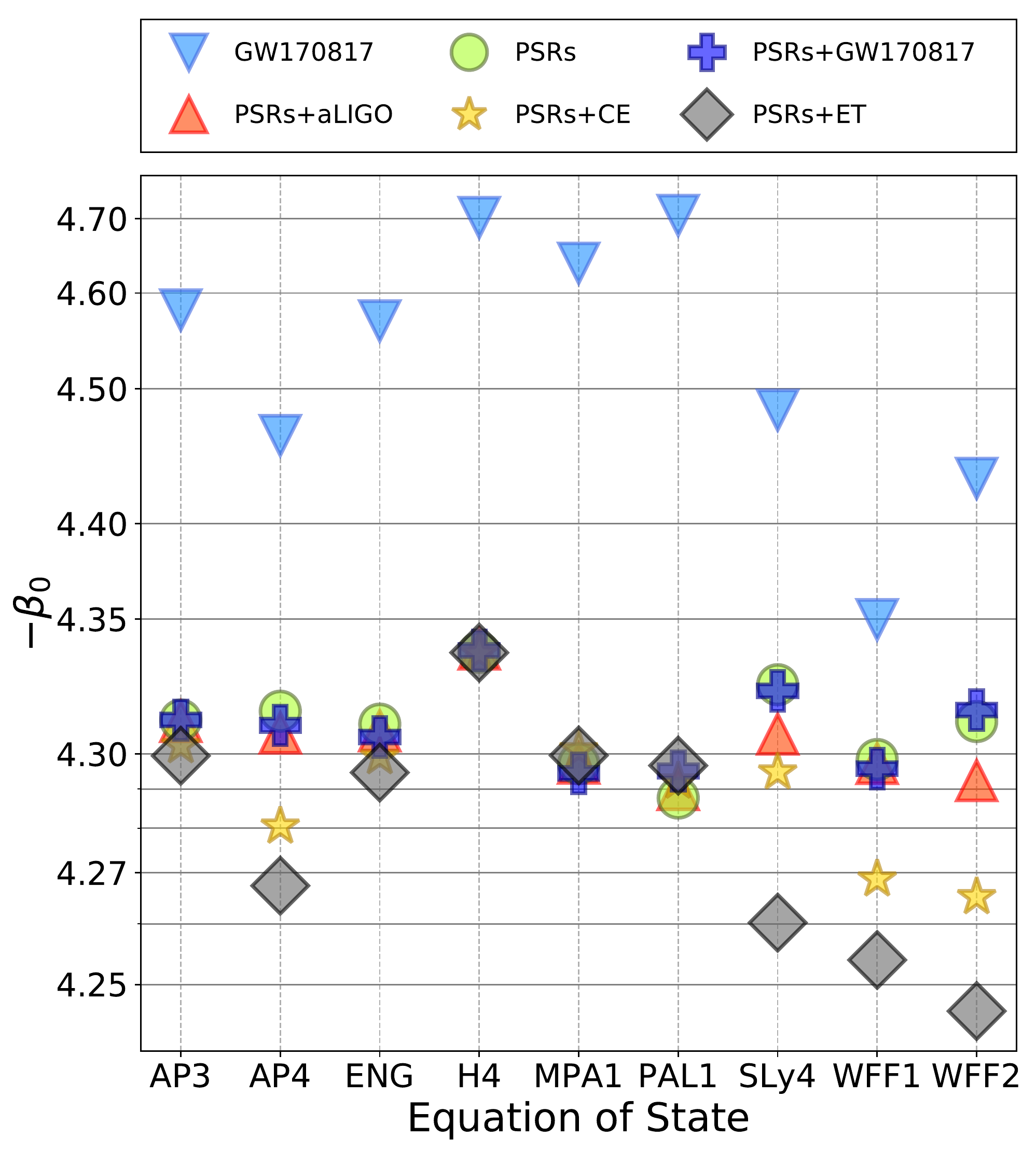}
\caption{(color online) Same as \cref{fig:alpha_90CL}, for the parameter,
$-\beta_0$. In order to express those constraints on $-\beta_0$ more clearly
around $-\beta_0 \sim 4.3$, a special logarithmic axis is adopted in this
figure.\label{fig:beta_90CL}}
\end{figure}
\begin{figure*}
  \includegraphics[width=16.5cm]{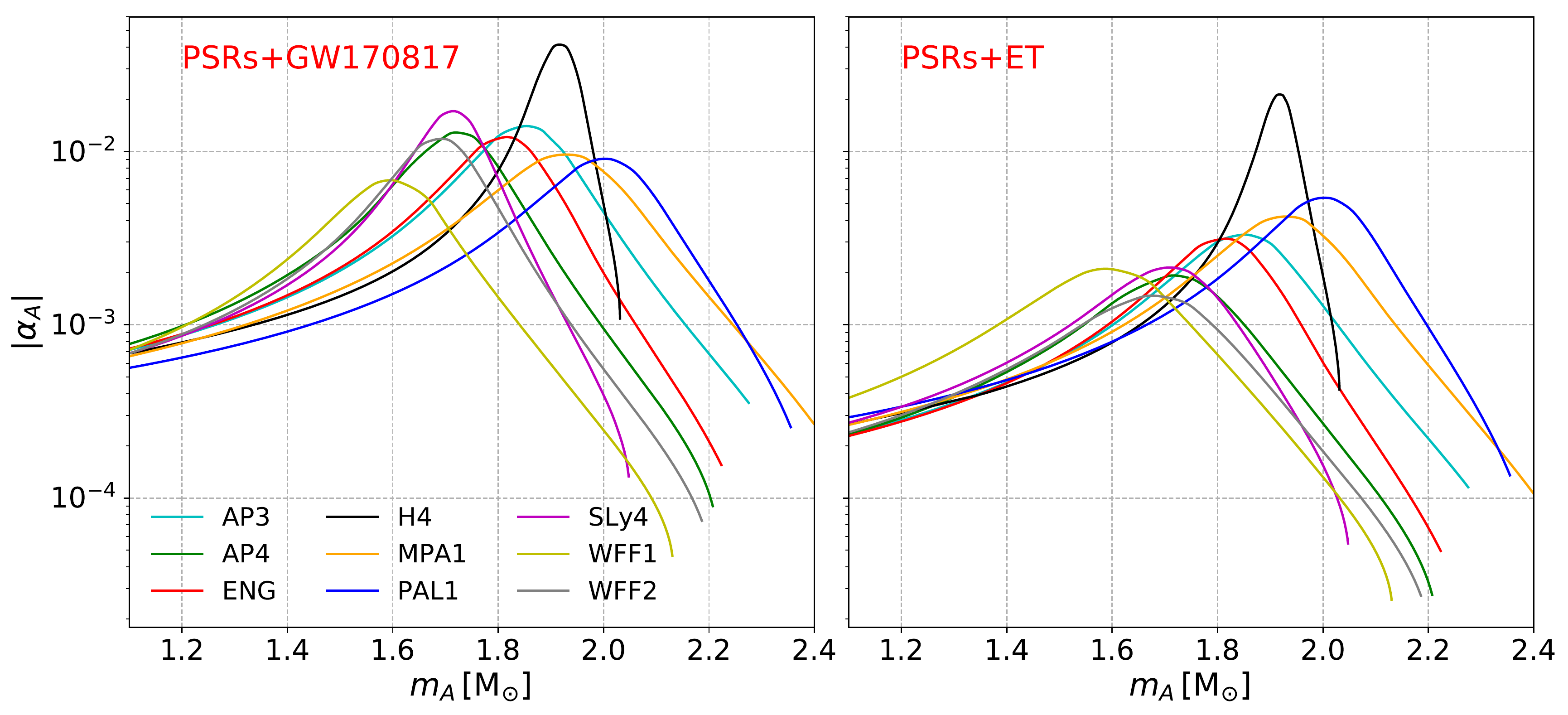}
\caption{(color online) The 90\% CL upper bounds on the NS effective scalar
coupling, $\alpha_A$, from the scenarios {\sf PSRs+GW170817} (left) and {\sf
PSRs+ET} (right). \label{fig:GWvsET90}}
\end{figure*}

In our study, we investigate all the scenarios mentioned above for nine EOSs.
The upper limits of the parameters, $\left| \alpha_0 \right|$, $-\beta_0$ at
90\% CL are illustrated in~\cref{fig:alpha_90CL,fig:beta_90CL} respectively.
\Cref{fig:app:GW,fig:app:Fisher} in~\cref{sec:app:mcmc} show the corresponding
marginalized 1-dimensional KDE distributions.

We collect our marginalized limits in \cref{fig:alpha_90CL,fig:beta_90CL}.
In~\cref{fig:alpha_90CL}, for nine EOSs we adopted, the scenario {\sf GW170817}
shows that $\left| \alpha_0 \right|$ cannot be well constrained.  With the five
binary pulsars, the constraints are improved enormously down to the level of $
\left| \alpha_0 \right| \lesssim 2 \times 10^{-4}$. It is evident that, the
scenarios, {\sf PSRs}, {\sf PSRs+GW170817}, and {\sf PSRs+aLIGO}, give similar
upper bounds on $\left| \alpha_0 \right|$. It dictates that, the constraints
from GW170817 are weaker than those from binary pulsars. The constraints on
$\left| \alpha_0 \right|$ are expected to be greatly enhanced with the
next-generation detectors, CE and ET.  Their upper bounds on $\left| \alpha_0
\right|$ can achieve the level of $\lesssim 1 \times 10^{-4}$.  Worthy to note
that, the constraints on $\left| \alpha_0 \right|$ given in
\cref{fig:alpha_90CL} are heavily influenced by our priors (especially the one
on $\beta_0$). A different prior will give a different limit, but the relative
strength of these observational scenarios is fixed.

In~\cref{fig:beta_90CL}, the bounds on $\beta_0$ from GW170817 are hardly
constrained for most of the EOSs. Only for the EOS {\sf WFF1}, $\beta_0$ can be
limited to a meaningful level, $-\beta_0 \lesssim 4.35$ at 90\% CL. In contrast,
for the scenarios, {\sf PSRs}, {\sf PSRs+GW170817}, and {\sf PSRs+aLIGO}, the
limits are improved up to the level of $\beta_0 \approx -4.30$ for most of the
EOSs (except for {\sf H4}). When considering the scenarios {\sf PSRs+CE} and
{\sf PSRs+ET}, we find that the bounds on $\beta_0$ have almost no improvement
for all EOSs, except for {\sf AP4, SLy4, WFF1} and {\sf WFF2}. The results of
$\beta_0$ are different from those of $\alpha_0$.  Particularly, the parameter
$\beta_0$ plays the major role in controlling the strength of scalarization. We
could not expect to obtain a tighter constraint on $\beta_0$ only by improving
the precision of observations. The results in~\cref{fig:beta_90CL} are
consistent with Fig.~7 in Ref.~\cite{Shao:2017gwu}.  In particular, for the EOSs
{\sf AP4, SLy4, WFF1} and {\sf WFF2}, CE and ET have the potential to improve
the current limit from binary pulsars.

We investigate the improvement from the scenario {\sf PSRs+ET}, with respect to
the scenario {\sf PSRs+GW170817}.  The upper bounds on the NS effective scalar
coupling $\alpha_A$, as a function of the NS mass $m_A$, are illustrated
in~\cref{fig:GWvsET90}. Here, the upper bounds at 90\% CL on theory parameters
(see \cref{fig:alpha_90CL,fig:beta_90CL}) are used for those curves. The area
below the curve for each EOS corresponds to the unconstrained region for a NS.
It is evident that, the allowed maximum effective scalar couplings are
constrained strongly for all EOSs except for the EOS {\sf H4}. The narrow range
of the NS mass, that corresponds to the scalarization peak, is around $m_A \sim
1.9\,\SUN$ for the EOS {\sf H4}. All NS masses from binary pulsars in
\cref{tab:PSR}, GW170817, and {\sf BNS@200Mpc} are not in this region. It is
also the reason why the constraints for the EOS {\sf H4} in~\cref{fig:beta_90CL}
are hardly improved, with the next-generation detectors.

Combining the results in~\cref{fig:GWvsET90,fig:beta_90CL}, we can understand
some results qualitatively. In the previous discussion, the upper bounds on
$\beta_0$ at 90\% CL can be placed at the level of $\sim -4.35$ for the EOS {\sf
WFF1}. It is the tightest constraint in the scenario {\sf GW170817}. Actually,
it can be understood in~\cref{fig:GWvsET90}. For the specific range of NS
masses, the spontaneous scalarization is significant. For the EOS {\sf WFF1},
the corresponding NS mass is around the value, $m_A \approx 1.4$--$1.7\,\SUN$.
In contrast, the other EOSs allow NSs to scalarize strongly when $m_A \gtrsim
1.7\,\SUN$. One of the BNS masses derived from GW170817 is in the range
$(1.36\,\SUN, 1.60\,\SUN)$. It is within the scalarization mass range for the
EOS {\sf WFF1}. Therefore, solely with GW170817, we can constrain $\beta_0$ for
this EOS.

In a short summary, the constraints on  $\left| \alpha_0 \right|$ improve with
the precision of the observations. But, for $\beta_0$, not only the precision of
the observations, but also the choice of the EOS can influence the limit.
Different EOSs allow NSs to scalarize at different NS masses. We can use the
observations, binary pulsars and BNSs, to constrain the DEF theory with
different EOSs, if suitable systems are observed.

%---------------------------------------------------------------------
\section{Conclusion}
\label{sec:conc}
%---------------------------------------------------------------------

In this paper, we studied the DEF theory in the nonperturbative strong field. In
the parameter space that we investigate, NSs could develop a phenomenon called
spontaneous scalarization. Instead of solving the modified TOV equations for NSs
(as was done in earlier study~\cite{Shao:2017gwu}), we have built  efficient
ROMs, implemented in the {\tt pySTGROM} package, to predict NS properties. The
code is made public for the community use. With the speedup of ROMs, MCMC
calculations were carried out to constrain the parameter space, $(\alpha_0,
\beta_0)$. Comparisons between various scenarios are discussed in detail.  We
summarize main points in the following.

\begin{enumerate}
    \item To speed up large-scale calculations, {\tt pySTGROM} is
    constructed for studying the DEF theory. Considering six
    scenarios that include currently available as well as future
    projected observations, we have tested our ROMs in practice. It
    turns out that, our ROMs are performed at least {\it two
    orders} of magnitude faster than the previous method. In the
    foreseeable future, we wish our ROMs to be used in relevant calculations
    for the DEF theory.
    \item For a fixed EOS, a NS is allowed to scalarize strongly in
    a specific mass range of the NS. A binary pulsar
    and/or a BNS with a particular NS mass in this mass
    region may be observed. They can be used to constrain the DEF
    theory stringently for this EOS. 
    \item Using the uncertainty in $\left| \Delta \alpha \right|$
    from the Fisher-matrix calculation~\cite{Will:1994fb,
    Shao:2017gwu}, the tightest constraints come from the scenario
    {\sf PSRs+ET} in our studies. Given our specific priors, we can
    bound the parameters to be $\left| \alpha_0 \right| \lesssim
    10^{-4.0}$ and $-\beta_0 \lesssim 4.3$ at  90\% CL. Due to
    its low-frequency sensitivity, ET for sure will provide us with
    significant improvement over current constraints on the dipole
    radiation.
    \item The spontaneous scalarization is mainly controlled by the
    $\beta_0$ parameter in the DEF theory. It is likely that, the
    constraints on the parameter $\left| \alpha_0 \right|$ can be
    constrained more tightly with the improvement in the observational
    precision. Different from $\left| \alpha_0 \right|$, the
    constraints on the scalarization parameter $\beta_0$ are
    related to the property of EOS, in addition to the precision of the
    observations.
\end{enumerate}

The {\it current} and {\it projected} bounds were obtained
in~\cref{ssec:PSRAndGW170817} and~\cref{ssec:future_measure} respectively. It
indicates that, the next-generation GW detectors, especially the ET, have the
potential to further improve current limits, set by the binary pulsars and
GW170817. Those current limits are to be improved over time, especially if
suitable systems filling the ``scalarization windows'' are
discovered~\cite{Shao:2017gwu}. Some binary pulsar systems, like
PSRs~J1012+5307~\cite{Lazaridis:2009kq} and J1913+1102~\cite{Lazarus:2016hfu},
still have large uncertainties in their masses. If the masses are constrained to
be around the ``scalarization window'', they may eliminate the possibility of a
strong scalarization below $2\,\SUN$. The new large radio telescopes, such as
the Five-hundred-meter Aperture Spherical radio Telescope (FAST) in
China~\cite{Nan:2011um,2016RaSc...51.1060L,Li:2019ads} and the Square Kilometre
Array (SKA) in Australia and South Africa~\cite{Kramer:2004hd, Shao:2014wja,
Bull:2018lat}, can help to improve the timing precision. In addition, they may
discover new systems that meet the requirements. On the other hand, O3 of
LIGO/Virgo detectors has began in April 2019. It has been discovering a handful
of GW candidates.\footnote{ GW candidates are collectively shown in the
gravitational-wave candidate event database (GraceDB) by the LIGO/Virgo
Collaboration in the link \url{https://gracedb.ligo.org/latest}.} By the end of
observing runs, some events may happen to be suitable systems to close the
scalarization windows. 
Asymmetric BNSs (with $\alpha_A \neq \alpha_B$) as well as NS-BHs (with
$\alpha_{\rm NS} \neq \alpha_{\rm BH} = 0$) would give asymmetric systems and
therefore could be particularly interesting.
Furthermore, the next-generation ground-based GW
detectors are expected to observe many more systems in the future, and they can
be used to study alternative gravity theories in a more precise way. Our ROMs
are built to meet the requirements of new observations to constrain the DEF
theory in an efficient yet accurate way.

%---------------------------------------------------------------------
\acknowledgments 
%---------------------------------------------------------------------

We are grateful to Bin Hu and Michael Kramer for comments.  We thank Norbert Wex
for stimulating discussions and carefully reading the manuscript.  This work was
supported by the Young Elite Scientists Sponsorship Program by the China
Association for Science and Technology (2018QNRC001), and was partially
supported by the National Natural Science Foundation of China (11721303,
11475006, 11690023, 11622546), the Strategic Priority Research Program of the
Chinese Academy of Sciences through the grant No. XDB23010200, the European
Research Council (ERC) for the ERC Synergy Grant BlackHoleCam under Contract No.
610058, and the High-performance Computing Platform of Peking University.  Z.C.
was supported by the ``Fundamental Research Funds for the Central
Universities''.

%---------------------------------------------------------------------
\appendix 
%---------------------------------------------------------------------

\section{The MCMC results for nine EOSs}
\label{sec:app:mcmc}

\begin{figure*}
  \centering
  \includegraphics[width=17.9cm]{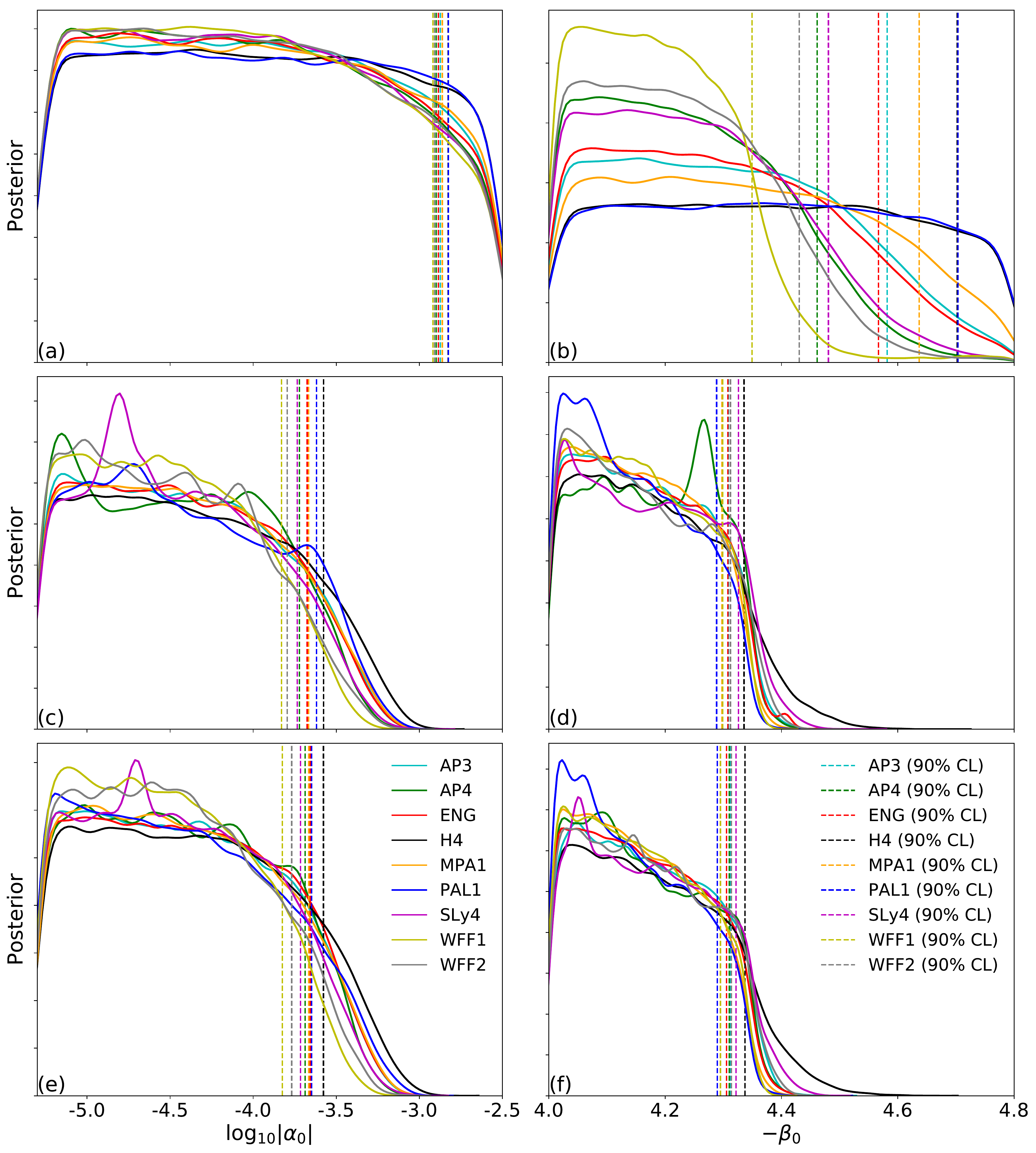}
\caption{(color online) The marginalized 1-dimensional KDE distributions for the
MCMC posteriors for the parameters $\log_{10} \left| \alpha_0 \right|$ and
$-\beta_0$. They are shown for nine EOSs and the scenarios, {\sf GW170817}
(top), {\sf PSRs} (middle) and {\sf PSRs+GW170817} (bottom). The distributions
of $\log_{10} \left| \alpha_0 \right|$ are shown in the left; the posterior
distributions of $- \beta_0$ are shown in the right. Their corresponding upper
bounds at 90\% CL are illustrated with dashed lines. \label{fig:app:GW}}
\end{figure*}
\begin{figure*}
  \centering
  \includegraphics[width=17.9cm]{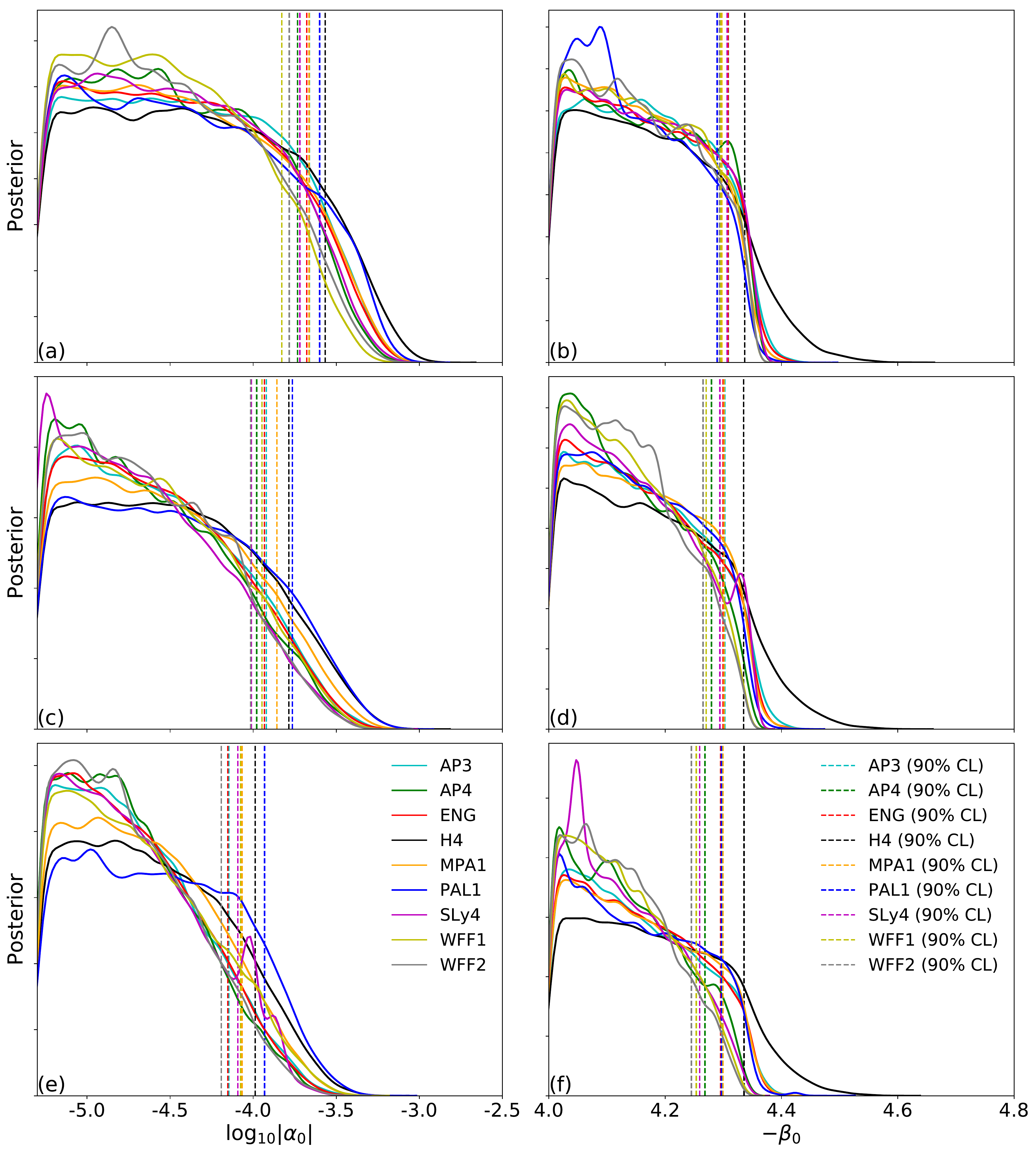}
\caption{(color online) Same as~\cref{fig:app:GW}, for the scenarios, {\sf
PSRs+aLIGO} (top), {\sf PSRs+CE} (middle) and {\sf PSRs+ET} (bottom).
\label{fig:app:Fisher}}
\end{figure*}

The marginalized 1-dimensional KDE distributions of the parameters, $\log_{10}
\left| \alpha_0 \right|$ and $-\beta_0$, for the EOSs in the set \{ {\sf AP3,
AP4, ENG, H4, MPA1, PAL1, SLy4, WFF1}, {\sf WFF2} \}, are illustrated
in~\cref{fig:app:GW,fig:app:Fisher}. Their corresponding upper bounds at 
90\% CL are shown with dashed lines.

%---------------------------------------------------------------------
\bibliography{refs}
%---------------------------------------------------------------------

\end{document}